\documentclass{article} 

\usepackage[utf8]{inputenc} 
\usepackage{hyperref}       
\usepackage{url}            
\usepackage{booktabs}       
\usepackage{graphicx}
\usepackage{xcolor}
\usepackage{listings}       
\usepackage{multirow}       
\graphicspath{ {./images/} }

\usepackage{arxiv}
\usepackage[numbers]{natbib}

\title{How the adoption of feature toggles correlates with branch merges and defects in open-source projects?}

\author{
  Eduardo S. Prutchi \\
  Instituto de Computação \\
  Universidade Federal Fluminense \\
  Niterói, RJ, Brazil \\
  \texttt{eduardosmil@id.uff.br} \\\And
  Heleno de S. Campos Junior \\
  Instituto de Computação \\
  Universidade Federal Fluminense \\
  Niterói, RJ, Brazil \\
  \texttt{helenocampos@id.uff.br} \\\And
  Leonardo G. P. Murta \\
  Instituto de Computação \\
  Universidade Federal Fluminense \\
  Niterói, RJ, Brazil \\
  \texttt{leomurta@ic.uff.br} \\
}

\begin{document}
\maketitle

\begin{center}
    Published version: \url{https://www.doi.org/10.1002/spe.3034}
\end{center}

\begin{abstract}
\textbf{Context}: Branching has been widely adopted in version control to enable collaborative software development. However, the isolation caused by branches may impose challenges on the upcoming merging process. Recently, companies like Google, Microsoft, Facebook, and Spotify, among others, have adopted trunk-based development together with feature toggles. This strategy enables collaboration without the need for isolation through branches, potentially reducing the merging challenges. However, the literature lacks evidence about the benefits and limitations of feature toggles to collaborative software development. \textbf{Objective/Method}: In this paper, we study the effects of applying feature toggles on 949 open-source projects written in 6 different programming languages. We first identified the moment in which each project adopted a feature toggles framework. Then, we observed whether the adoption implied significant changes in the frequency or complexity of branch merges and the number of defects, and the average time to fix them. Finally, we compared the obtained results with results obtained from a set of control projects that do not use feature toggles frameworks. \textbf{Results/Conclusion}: We could observe a reduction in the average merge effort and an increase in the average total time needed to fix defects after adopting feature toggles frameworks. However, we could not confirm that this increase was influenced by the use of feature toggles.
\end{abstract}


\section{Introduction}\label{sec:introduction}
Branching techniques, supported by version control systems, have been widely adopted to leverage collaborative software development. On the one hand, it allows temporary code isolation, favoring parallel development. On the other hand, it may demand an additional effort to merge the code developed in parallel back to the mainline of development. Depending on many factors, such as isolation and code complexity, branching can bring risks to the project due to time-consuming and error-prone merges \cite{shihab_effect_2012_1}. 

Recently, leading global companies like Google, Microsoft, Facebook, and Spotify, among others, have adopted feature toggles as an alternative to branches for the development of their products. This technique allows a more seamless collaborative development process, where new features are implemented directly in the mainline of development (i.e., trunk-based development) without the need for creating branches. As a natural consequence, developers expect to get rid of complicated branch merges \cite{feitelson_development_2013_2, heys_alm_nodate_3, rahman_feature_2016_4, kumar_dilemma_2017_5, rahman_modular_2018_6}. 

Although the use of feature toggles is increasing in companies, the literature lacks scientific evidence about the benefits and limitations of using it for collaborative software development. To the best of our knowledge, Meinicke et al.~\cite{meinicke2020capture} is the only available empirical study about feature toggles over a range of open-source projects. Besides, only a few existing works discuss experiences of using feature toggles based on specific case studies or surveys. For instance, Rahman et al.~\cite{rahman_feature_2016_4} report the changes in collaborative development after the feature toggles introduction in Google Chrome and what was necessary to control the toggles debt. Additionally, Schermann et al.~\cite{schermann_empirical_2016_7} present an overview of continuous delivery practices based on a literature survey, including the use of feature toggles. Neely and Stolt~\cite{neely_continuous_2013_8} discuss the use of the feature toggle technique instead of branching over a continuous delivery process. Finally, Rehn~\cite{rehn_continuous_2012_9} compared some collaborative development techniques and suggested the use of feature toggles for continuous integration instead of feature branches. Nevertheless, none of them provides quantitative evidence, based on a large project corpus, about the benefits and limitations of using feature toggles instead of branching for collaborative software development. Additionally, we could not find any study about the effects of using feature toggles on the number and time to fix defects.

The goal of this paper is three-fold: (1) analyze how widespread is the adoption of feature toggles frameworks in open-source projects; (2) analyze whether the use of feature toggles frameworks implies changes in the frequency or complexity of branch merges; and (3) analyze whether the use of feature toggles frameworks is associated with changes on the number and fixing times of defects in open-source projects. To achieve our goal, we considered a corpus of 949 open-source projects written in 6 different programming languages. We first identified the moment that each project adopted a feature toggles framework and evaluated whether the adoption implied changes in the frequency or complexity of the merges. We also evaluated if the frequency or the average time to fix defects was affected. Finally, we compared the obtained results with results obtained from a set of control projects that do not use feature toggles frameworks.

In general, we could not observe a statistically significant difference in the frequency of branch merges. However, we could observe a significant decrease in the merge effort after adopting feature toggles frameworks. Furthermore, although the average number of defects and the average total time spent fixing them increased, we could not observe a statistically significant difference. 

This paper is organized into six other sections. In Section~\ref{sec:background}, we introduce the concepts of feature toggles. In Section~\ref{sec:materials}, we formulate our research questions and describe the methods adopted in our study to answer them. In Section~\ref{sec:results}, we present the results of our study, together with discussions. We also discuss, in  Section~\ref{sec:threats}, the threats to the validity of this study. The related works are presented in Section~\ref{sec:related}. Finally, Section~\ref{sec:conclusions} presents the conclusion and highlights some future work.

\section{Background}\label{sec:background}

Feature Toggles, also known as Feature Flags, Feature Flippers, or Feature Switches, consist of surrounding features (functionalities) in the code with if statements to gain more control over their release process. By surrounding a feature with a toggle (if statement), developers can decide when and for whom the feature should be available. It means that feature toggles allow creating dynamic behavior flows without the need to recompile the code and subsequent deployments. 

\begin{table}
\caption{Simple Implementation of a feature toggle in C\#.}
\label{tbl:tbl1}
\begin{center}
\begin{tabular}{ll}
\toprule
Before using feature toggle & After using feature toggle \\
\midrule 
\begin{lstlisting}
static void main() {
    ShowShoppingCart();
}

public bool ShowShoppingCart() {
    ...
}
%*\newline*)
%*\newline*)
%*\newline*)
%*\newline*)
\end{lstlisting}
&
\begin{lstlisting}
static void main() {
    if (getFeatureIsEnabled("useNewShoppingCart"))
        ShowNewShoppingCart();
    else
        ShowShoppingCart();
}

public bool ShowNewShoppingCart() {
    ...
}

public bool ShowShoppingCart() {
    ...
}
\end{lstlisting} \\
\bottomrule 
\end{tabular}
\end{center}
\end{table}

Table~\ref{tbl:tbl1} shows a simple implementation of a feature toggle in C\# regarding a shopping cart method in e-commerce software. In this example, a team wants to develop a new shopping cart (ShowNewShoppingCart) and, at the same time, continue using the current shopping cart (ShowShoppingCart). On the left-hand side, Table~\ref{tbl:tbl1} shows the code before using the feature toggle. On the right-hand side, the code contains a feature toggle. Thus, this new feature will be enabled only for the developer team and disabled for other users.

Therefore, feature toggles enable Dark Launching, which consists of releasing disabled partial features directly in the production environment \cite{neely_continuous_2013_8, parnin2017adages}. This specific property means that developers may change code directly in the main development line (i.e., trunk-based development) of the project instead of creating feature branches \cite{berczuk_pragmatic_2003_10} for isolating parallel changes. Thus, reducing the isolation potentially decreases the explicit merging needs. For instance, a team could add a new feature directly to the mainline, keeping it disabled by using a specific toggle while it is under development. Even if the code is released into production, it will not be available to users due to the toggle. When the feature is ready and tested, it can be released by merely switching the toggle or removing the toggle from the code.

Feature toggles are also useful for supporting other techniques, such as canary releases and A/B testing. According to Sato~\cite{sato_bliki_2014_11}, canary releases aim at reducing the risk of introducing a new feature into production. This technique allows for rolling out the feature to a specific group of users, selected by region or other characteristics. These users act as beta testers over the new feature. Complementarily, A/B testing enables releasing a specific feature to a group of users and another feature to another group. By observing a dependent variable, one can run hypothesis tests and identify the feature that is best suited to the task. 

Many companies have focused on reducing the risks associated with software integration (i.e., merging code) by adopting continuous integration \cite{digitalocean_developers_2018_12}. Rahman et al.~\cite{rahman_feature_2016_4} suggest the use of the feature toggles technique together with the continuous integration process. Feature toggles allow the continuous integration infrastructure to either test the current implementation of a feature or ignore it by enabling or disabling the toggle, respectively.

According to Fowler~\cite{fowler_bliki_2010_13} and Hodgson~\cite{pete_hodgson_feature_2016_14}, feature toggles can be divided into two types: release and business. Release (or development) toggles are usually temporary and will be disabled by default during the development of a feature, being switched on just for test purposes. After having the feature implemented and tested, the toggle could be removed from the code. Fowler emphasizes that managing the toggles is always essential, particularly for release toggles, by removing those that already have bedded down into production. On the other hand, business or long-term toggles encapsulate features that can be managed by end-users, being enabled or not, depending on the user needs or product configuration \cite{hubaux_unifying_2012_15}. For instance, such kinds of toggles can be managed by end-users of Google Chrome using the “chrome://flags/” URL.

\section{Materials and Methods}\label{sec:materials}

The literature has discussed in general the possible benefits of using feature toggles in software projects. However, in this paper, we focus on studying if the use of feature toggles correlates with the number and effort of branch merges, and with the number and time to fix defects. To reach this objective, we considered a corpus of projects that adopted some feature toggle framework in a specific moment of their history. Then, we contrasted the history before and after the adoption of feature toggles.

In this section, we present our research questions, the frameworks used to identify projects that use feature toggles, and how we constructed and filtered our project corpus.

\subsection{Research Questions}\label{sec:rqs}

We have elaborated three research questions that guide our study. We describe each question here and detail them further in Section~\ref{sec:results} together with their answers.

\textit{RQ1: What is the adoption level of feature toggles in open-source projects?}

This question investigates the distribution of feature toggle frameworks being adopted by the open-source community. Moreover, we characterize the corpus of projects that use feature toggles, presenting which programming languages are primarily used, the commit count, and the exact moment they adopted feature toggles.

\textit{RQ2: Do the number of branch merges and the necessary effort change after adopting feature toggles?}

Feature toggles are believed to help keep branches short-lived, by enabling trunk-based development and thus, reducing the difficulty of performing merges \cite{pete_hodgson_feature_2016_14, adams2016modern}. This is due to the fact that long-living branches are often associated with merge conflicts \cite{dias2020understanding}, making the merge task more painful for the developers \cite{phillips2011branching}. Despite this belief, to the best of our knowledge, there is no current empirical evidence that the use of feature toggles impacts on the number and the effort required for performing merges. Based on this, and considering the exact moment of adoption of feature toggles in each project, we investigate whether the adoption of feature toggles is associated with the number of merges and their required effort. 

We split this research question into three other specific sub-questions (RQ2.1, RQ2.2, and RQ2.3). In RQ2.1, we evaluate whether the number of branch merges changes after adopting feature toggles. We also evaluate whether this behavior is the same across different programming languages. In RQ2.2, we analyze whether the branch merge effort changes after the adoption of feature toggles. Finally, in RQ2.3, we investigate whether the total branch merge effort changes after adopting feature toggles. This last research question combines RQ2.1 and RQ2.2 by considering both the number of branch merges and their respective effort.

As the number of commits before and after the adoption of feature toggles may be a confounding factor for this research question, we decided to normalize our data in terms of the number of commits. A natural normalization would be ``merges per commit", but this would lead to small decimal numbers, as merges are less frequent than commits. Consequently, we opted to use ``merges per 100 commits". After this normalization, we have the following dependent variables: normalized number of merges (i.e., (\# merges x 100) / \# commits) for RQ2.1, effort per merge (i.e., merge effort sum / \# merges) for RQ2.2, and normalized merge effort (i.e., normalized number of merges x effort per merge) for RQ2.3.

\textit{RQ3: Do the number of software defects and the time to fix them change after adopting feature toggles?}

According to Fowler~\cite{fowler_bliki_2010_13}, feature toggles bring a challenge to software testing due to the number of different toggle combinations. However, Fowler~\cite{fowler_bliki_2010_13} mentions that only two types of combinations need to be tested: all the toggles that are expected to be “on” in the next release and all toggles together. Therefore, considering the testing complexity in this scenario and that the lack of testing may be associated with defects, it remains an open question whether feature toggles usage is associated with defects. Thus, in this research question, we evaluate if the number of defects and time to fix them change after adopting feature toggles.

Therefore, similarly to RQ2, we also split this research question into three sub-questions (RQ3.1, RQ3.2, and RQ3.3). In RQ3.1, we evaluate whether the number of defects changes after adopting feature toggles. In RQ3.2, we analyze whether the time needed for fixing a defect changes after the adoption of feature toggles. Finally, in RQ3.3, we analyze whether the total time spent fixing defects changes after adopting feature toggles. Again, RQ3.3 analyzes the results of RQ3.1 and RQ3.2 combined, as it considers both the number and the duration of defects.

For this research question, not only the number of commits before and after the adoption of feature toggles may be a confounding factor, but also the size – the larger the project, the higher the absolute number of defects. For this reason, we decided to normalize our data in terms of both the number of commits and the number of lines of code. After this normalization, we have the following dependent variables: normalized number of defects (i.e., (\# defects x 100) / \# commits / (KLOC / \# commits)) for RQ3.1, time per defect for RQ3.2 (i.e., defects time sum / \# defects), and normalized time fixing defects (i.e., (defects time sum x 100) / \# commits / (KLOC / \# commits)) for RQ3.3.

\subsection{Feature Toggles Frameworks}

A feature toggle can be a simple conditional statement (if-then-else) that is responsible for defining an execution flow in a software application. Thus, identifying them in the source code is not trivial. On the other hand, some frameworks  enable the use of feature toggles on software projects. In this study, we focus on projects that use such frameworks. Thus, we first carried out a study to identify existing open-source frameworks that support feature toggles.

Unfortunately, we could not find any reliable material that lists existing feature toggles frameworks. Thus, we opted to perform an extensive search of feature toggles frameworks on the Internet. Initially, we searched for frameworks on reference websites about feature toggles (\url{http://enterprisedevops.org} and \url{http://featureflags.io}) and also on DevOps books \cite{httermann_devops_2012_16}. Besides, we mined repositories citing “feature toggle framework” (or feature flags, feature switches) in their description. We could select feature toggle frameworks for six different programming languages.

Next, for each framework, we read their documentation and examples to identify keywords or code snippets that indicate whether a project adopts the framework. Most of the identified code snippets are class imports. Table~\ref{tbl:tbl2} shows the feature toggle frameworks identified by our study, grouped by their respective programming language. We also show in this table, the keyword used to identify whether a project has instantiated the framework. 

\begin{table}
\caption{Feature toggles frameworks considered in our study, together with the keywords used to indicate whether a project has instantiated the framework. Keywords are displayed as regular expressions similar to how they appear in the frameworks documentation. The search process is case-insensitive. Some of the keywords were tuned to reduce false positive results for some frameworks (e.g. by including a semicolon or using partial import snippets).}
\label{tbl:tbl2}
\begin{center}
\begin{tabular}{ccl}
\toprule
Prog. Lang.                 & Framework                       & Keywords                                             \\ \toprule
\multirow{4}{*}{C\#}        & Switcheroo                      & IFeatureToggle                                       \\ \cline{2-3} \rule{0pt}{2ex}
                            & FeatureSwitcher                 & using FeatureSwitcher                                \\ \cline{2-3} \rule{0pt}{2ex}
                            & \multirow{2}{*}{FeatureToggle}  & using FeatureToggle                                 \\ \cline{3-3} \rule{0pt}{2ex}
                            &                                 & using FeatureToggle.Toggles;                         \\ \hline \rule{0pt}{2ex}
\multirow{2}{*}{Java}       & Togglz                          & import org.togglz.core.feature                       \\ \cline{2-3} \rule{0pt}{2ex}
                            & FF4J                            & import org.ff4j.FF4j                                 \\ \hline \rule{0pt}{2ex}
\multirow{4}{*}{JavaScript} & Ericelliot/feature-toggle       & require(``feature-toggles'')                        \\ \cline{2-3} \rule{0pt}{2ex}
                            & angular-toggle-switch           & module.provider\textbackslash{}(`toggleSwitchConfig' \\ \cline{2-3}  \rule{0pt}{2ex}
                            & fflip                & require(.fflip.)                                    \\ \cline{2-3} \rule{0pt}{2ex}
                            & ember-feature-flags             & config.featureFlags                                  \\ \hline \rule{0pt}{2ex}
PHP                         & Qandidate Toggle & Qandidate\textbackslash{}Toggle                                     \\ \hline \rule{0pt}{2ex}
\multirow{6}{*}{Python}     & Gutter                          & gutter.client                                        \\ \cline{2-3} \rule{0pt}{2ex}
                            & Gargoyle                        & from gargoyle import gargoyle                        \\ \cline{2-3} \rule{0pt}{2ex}
                            & \multirow{3}{*}{django-waffle}  & waffle.decorators                                    \\ \cline{3-3} \rule{0pt}{2ex}
                            &                                 & from waffle                                          \\ \cline{3-3} \rule{0pt}{2ex}
                            &                                 & import waffle                                          \\ \cline{2-3} \rule{0pt}{2ex}
                            & Flask-FeatureFlags              & from flask\_featureflags                             \\ \hline \rule{0pt}{2ex}
\multirow{4}{*}{Ruby}       & Rollout                         & \$rollout = Rollout.new(\$redis)                       \\ \cline{2-3} \rule{0pt}{2ex}
                            & \multirow{2}{*}{featureflags}   & Featureflags.defaults                                \\ \cline{3-3} \rule{0pt}{2ex}
                            &                                 & class Admin::FeaturesController                      \\ \cline{2-3} \rule{0pt}{2ex}
                            & feature\_flipper                & FeatureFlipper.features do                           \\ \bottomrule
\end{tabular}
\end{center}
\end{table}

\subsection{Project Corpus}\label{sec:corpus}

We accessed the GitHub API v3 and queried\footnote{\url{https://docs.github.com/en/rest/reference/search\#search-code}} for projects with any of the keywords shown in Table~\ref{tbl:tbl2}. The initial corpus ($C_{Initial}$) was composed of 1,001 projects, comprising six different programming languages. After a first analysis of the corpus, we realized that some projects were not instances, but the framework itself. Some other projects are forks from another project in the corpus. Thus, $C_{RQ1}$ was formed after removing these projects. It includes 949 projects, comprising six different programming languages, as shown in Table~\ref{tbl:tbl3}. For every programming language, the standard deviation of the number of commits is much higher than the respective mean. It means that there is a spreading of the number of commits on the projects. Moreover, some projects have just one or few commits, not being relevant to our study. They were filtered out, as explained in Section~\ref{sec:filtering}. We can also observe that most projects in $C_{RQ1}$ were written in JavaScript, compared to other programming languages. 

\begin{table}
\caption{Distribution of projects in $C_{RQ1}$.}
\label{tbl:tbl3}
\begin{center}
\begin{tabular}{lrrrrrrrr}
\toprule
\multirow{2}{*}{\begin{tabular}[c]{@{}l@{}}Prog.\\    Lang\end{tabular}} & \multirow{2}{*}{\# Proj} & \multicolumn{7}{c}{\# Commits}         \\ \cline{3-9} 
                                                                        &                          & Mean & Std. Dev.  & Min & Q1 & Med & Q3   & Max    \\
\toprule
C\#                                                                     & 85                       & 337 & 1,022 & 1   & 6  & 22  & 59   & 5,196  \\
Java                                                                    & 197                      & 791 & 2,157 & 1   & 3  & 15  & 119  & 18,035 \\
JavaScript                                                              & 373                      & 411 & 3,283 & 1   & 3  & 13  & 84   & 44,228 \\
PHP                                                                     & 14                       & 411 & 751 & 1   & 11 & 50  & 210  & 2,071  \\
Python                                                                  & 151                      & 3,438 & 10,118 & 1   & 11 & 112 & 1691 & 59,712 \\
Ruby                                                                    & 129                      & 792 & 2,528 & 1   & 9  & 60  & 792  & 19,847 \\
\midrule
Total                                                                   & 949                      &    &   &     &    &     &      &       \\
\bottomrule
\end{tabular}
\end{center}
\end{table}

We cloned all 949 projects in May/2018 and automatically analyzed each project’s history to segregate commits that occurred before and after the introduction of the feature toggles framework.To identify the commit where the feature toggles framework was introduced, we used the \textit{git bisect}\footnote{\url{https://git-scm.com/docs/git-bisect}} in combination with the \textit{git grep}\footnote{\url{https://git-scm.com/docs/git-grep}} command, using the keywords for the respective framework on each cloned project. The \textit{git bisect} command enables the use of a binary search in the history of each project. In addition, the \textit{git grep} command enables a regular expression search on all files for a given commit. After identifying the exact commit where the feature toggles framework was introduced in each project, we extracted the following information regarding the history before and after the adoption of the framework: number of commits, number of developers, description, dates, and labels of issues/pull requests (open and closed), and number of branch merges, together with their respective effort, as explained in the following. 

\subsubsection{Branch merges}

Our interest in this paper is on studying branch merges instead of workspace merges, which occur as a natural consequence of concurrent development. Workspace merges are usually due to short-term unnamed branches created by the clone operation. They integrate contributions of just one developer, and this developer is in charge of performing the merge. Conversely, branch merges are usually long-term and involve multiple developers \cite{costa_characterizing_2014_17}, being harder to perform and, consequently, dreaded by developers. We adopted the heuristic proposed by Costa et al.~\cite{costa_tipmerge_2016_18} to identify whether a merge commit is due to a named branch. This heuristic considers a merge commit as a branch merge if more than one unique developer has contributed to each side of the merge. Additionally, we analyzed the log message of every merge commit, searching for the expression “merge branch”. From now on, all mentions to merge refers to branch merges, unless specified otherwise.

\begin{figure}
    \centering
    \includegraphics[width=0.8\textwidth]{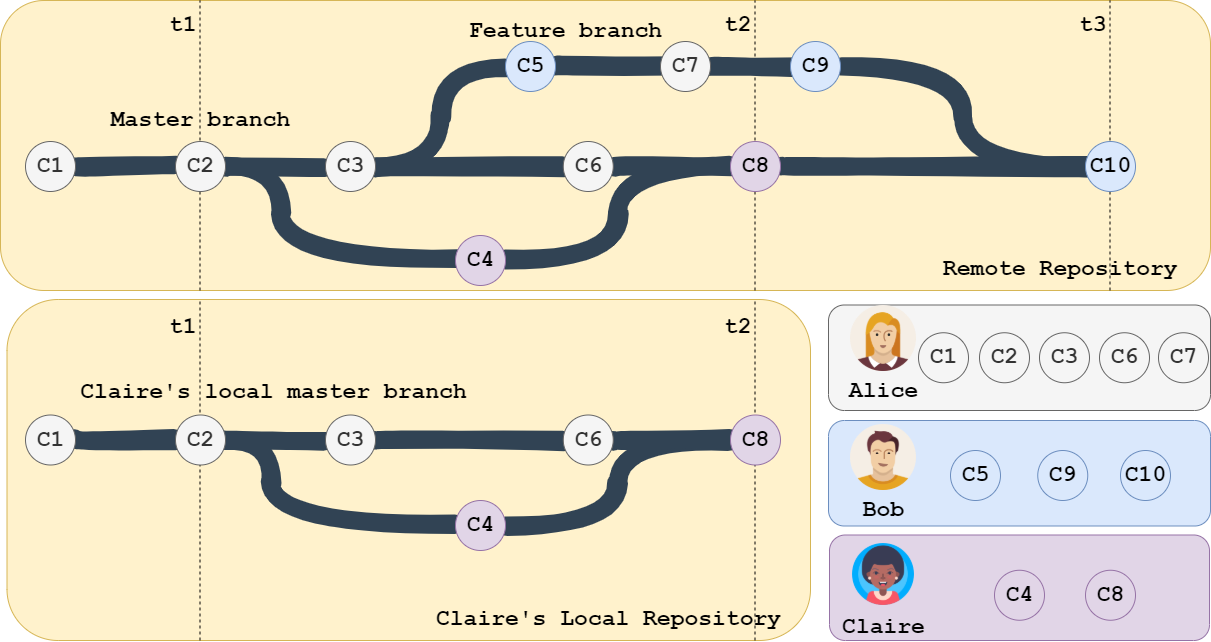}
    \caption{Development scenario illustrating a workspace merge (C8) and a branch merge (C10).}
    \label{fig:merges_scenario}
\end{figure}

Figure~\ref{fig:merges_scenario} illustrates a scenario where a branch merge and a workspace merge occur. Consider a development team composed of Alice, Bob, and Claire. In this scenario, Alice is the main developer in charge of the master branch, while Bob and Claire are developing new features. For simplicity's sake, Figure~\ref{fig:merges_scenario} shows only the remote repository and Claire’s local repository. Moreover, \textit{t1} represents the moment when Claire pulled the remote master branch to her local repository to start working on a feature. She made some changes and included them in C4. Later, on \textit{t2}, she wanted to push C4 to the remote repository. However, the remote master branch already had changes (C3 and C6) that she did not have. Hence, she needed first to pull these changes to her local repository, perform a merge with her local changes and then push her local commits (C4 and C8) back to the remote repository. The merge commit C8 is considered a workspace merge since only one developer contributed to a branch (Claire). Bob, who has been working on the feature branch together with Alice, authored the C10 merge commit in \textit{t3}. In this case, one side of the C10 merge commit has contributions from both Alice (C7) and Bob (C5 and C9), and the other side has contributions from Claire (C8) and Alice (C6). Thus, C10 is classified as a branch merge.

\subsubsection{Merge effort}

To investigate the effects of feature toggles on the merge effort, we adopted a metric defined by Prudêncio et al. \cite{prudencio_lock_2012_19} and implemented by Moura and Murta \cite{moura_uma_2018_20}. This metric uses diff operations on both files being merged, their common ancestor, and the merged version to estimate the actions performed during the merge. These actions are represented by the number of added and removed lines of code (code churn). The actions are used to calculate the merge effort (i.e., existing actions in the merged version but not in its parent versions) performed by the developers to integrate the versions. Hence, if the merge was performed automatically by the Version Control System (VCS), no merge effort was needed from the developers. Otherwise, they had to manually resolve existing conflicts. In this paper, we use this merge effort as a proxy for the effort required by the developers during a merge.

A more technical explanation of how we compute the merge effort can be found in Moura and Murta~\cite{moura_uma_2018_20}, but we summarize it in the following. First, we identify the code churn of each branch by performing a diff between the base version (i.e., the common ancestor) and the tip of the branch. These two sets of actions (lines of code added and removed in the branches) are combined, producing a multiset \cite{artofprogramming} with all actions performed in the branches. Then, we identify the code churn of the merge by performing a diff between the base version and the merge version. This produces a multiset with all actions that were actually committed in the merge. Finally, we compute the merge effort by subtracting the former multiset from the latter. The produced multiset contains just the lines of code added or removed for resolving the merge conflicts, and the merge effort is the total number of actions in this multiset. 

For instance, a merge that combines two independent methods added in separate files would lead to zero merge effort, since the VCS would perform it automatically. Similarly, if these two independent methods are added to the same file, but in different regions, the merge effort would also be zero, since no additional actions would be needed to conciliate the branches. However, the integration of a new feature implemented in parallel to an extensive refactoring would lead to a significant merge effort to adjust the feature to the new code organization imposed by the refactoring.

\subsubsection{Defect issues}

It is not trivial to obtain information regarding software defects in general. GitHub offers a mechanism in their website to manage issues. They can be related to documentation, instructions about how to use the project, reporting a bug, seeking contributors, etc. In this paper, we decided to use GitHub issues that are related to defects in the projects. To determine whether a project issue is associated with a defect or not, we checked whether the issue labels contain one of the following words: “Bug”, “kind/bug”, “Priority: Critical”, “Priority: Medium”, “Type – Bug”, “install-bug”, “404”, “403”, “type: bug”, “bug (open source)”, “error”, “contrib: good first bug”, “contrib: maybe good first bug”, “hotfix”, “incorrect”, and “mistake”. As this criteria lead to many false negatives, we also searched over the title and the description of the issue for the following words: “fix”, “error”, “problem”, “invalid”, “defect”, “500”, “404”, “403”, “exception”,  “bug”, “resolve”, “does not”, “exception thrown”, “not able”, “hotfix”, “incorrect”, “mistake”, “broken”, “not work”, “not respond”, “unable to”, “failing”, “failure”, “502”, “cannot”, “troubleshooting”, and “wrong”.

The next task was to define whether the defect issue occurred before or after the adoption of feature toggles. Unfortunately, $C_{RQ1}$ has few projects that link commits to issues. Alternatively, we contrasted the issue creation/closing date with the date when the feature toggles framework was adopted, leading to the following scenarios: (1) if the issue was closed before the adoption of feature toggles, then it was classified as “before feature toggles”; (2) if the issue was created after adoption of feature toggles, then it was classified as “after feature toggles”; and (3) if the issue was created before and closed after the adoption of feature toggles, then it was discarded.

\subsection{Corpus Filtering}\label{sec:filtering}

\begin{figure}
  \centering
  \caption{Corpora used to answer each research question.}
  \label{fig:fig1}
    \includegraphics[trim=0cm 18cm 0cm 2cm, width=0.8\textwidth]{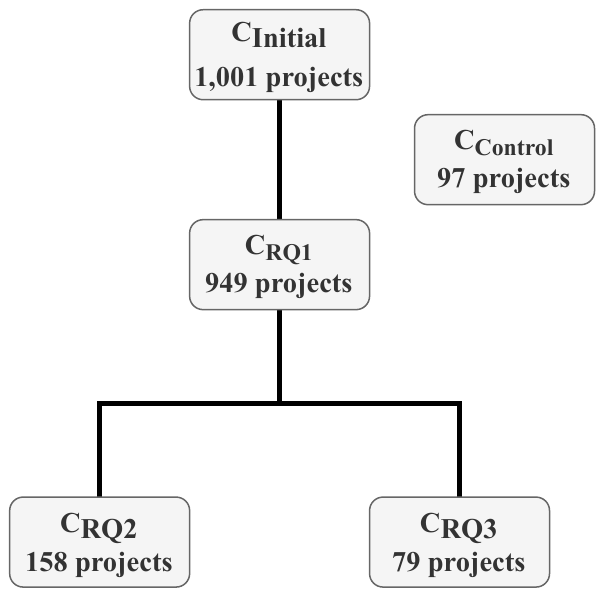}
\end{figure}

In an initial analysis, we found multiple repositories in the corpus with a very low level of commits. These repositories would not be appropriate for answering RQ2 and RQ3. Thus, we filtered the corpus again to respect the specific needs of these two research questions, as discussed in the remaining of this section. Figure~\ref{fig:fig1} shows the project corpora used for answering each research question, their derivation, and their size in terms of number of projects. Note that we used different corpora for RQ2 and RQ3, but both were derived from $C_{RQ1}$. The $C_{Control}$ corpus will be discussed in the next section.

\subsubsection{Corpus for RQ2}

As previously discussed, RQ2 contrasts the density of merge commits before and after introducing the feature toggles framework. Consequently, the corpus for such question should only contain projects that have enough commits (before and after) that allow at least one merge commit. This leads us to an intermediate question: what is the minimum number of commits in a project of $C_{RQ1}$ for having at least one merge?

\begin{figure}
  \centering
  \caption{Distribution of commits per merge in projects with at least one merge in $C_{RQ1}$. The boxplot shows that all projects with at least 82 commits have at least one merge commit, except for outliers.}
  \label{fig:fig2}
\includegraphics[width=0.7\textwidth]{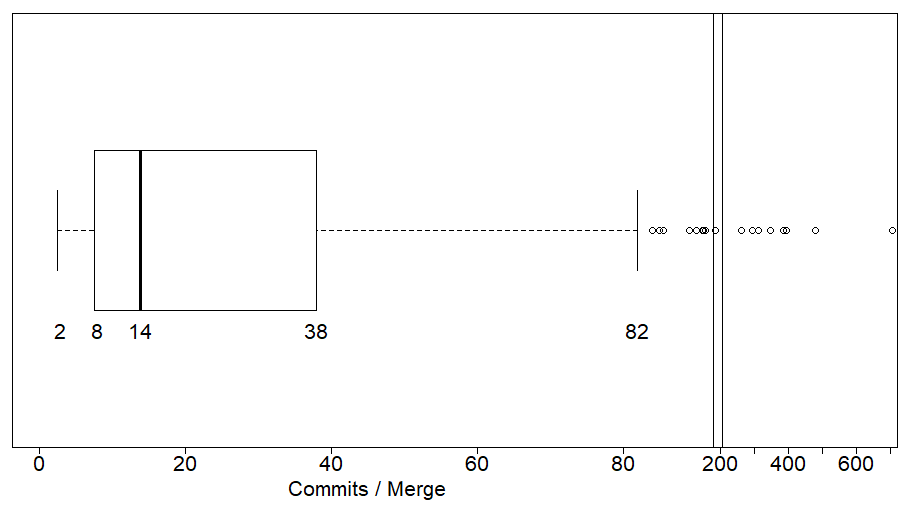}
\end{figure}

Therefore, aiming at finding such commit threshold, we first selected all projects that have at least one merge commit in $C_{RQ1}$. This sub-set is composed of 414 projects. Using this sub-set, we computed the distribution of commits per merge for each project, as shown in Figure~\ref{fig:fig2}. The upper limit of this boxplot, calculated using Tukey's fences formula $Q3 + 1.5 \times IQR$ \cite{barnett_outliers_1994_21}, is 82 commits per merge. The interpretation of such threshold is that all projects with at least 82 commits have at least 1 merge commit, except outliers.

We used this threshold to select projects with at least one merge commit that have at least 82 commits before and 82 commits after the introduction of the framework. This guarantees that any project in the corpus has enough commits before and after introducing feature toggles to have merges. It also guarantees that any selected project has enough commits to be representative. Thus, a new corpus was created for RQ2, named $C_{RQ2}$, with such projects. It contains 158 projects, still covering six programming languages, as characterized in Table ~\ref{tbl:tbl4}.

\begin{table}
\caption{Distribution of the projects in $C_{RQ2}$}.
\label{tbl:tbl4}
\begin{center}
\begin{tabular}{lrrrrrrr}
\toprule
\multirow{2}{*}{\begin{tabular}[c]{@{}l@{}}Prog\\    Lang\end{tabular}} & \multirow{2}{*}{\# Proj} & \multicolumn{6}{c}{\# Commits}                 \\ \cline{3-8} 
                                                                        &                          & Mean  & Min   & Q1    & Med   & Q3    & Max    \\
\toprule
C\#                                                                     & 8                        & 3,200 & 316   & 2,888 & 3,504 & 3,742 & 5,196  \\
Java                                                                    & 37                       & 3,990 & 296   & 1,329 & 2,831 & 6,127 & 18,035 \\
JavaScript                                                              & 26                       & 4,582 & 240   & 385   & 708   & 2,663 & 44,228 \\
PHP                                                                     & 3                        & 1,731 & 1,086 & 1,562 & 2,037 & 2,054 & 2,071  \\
Python                                                                  & 55                       & 9,191 & 387   & 1,512 & 2,325 & 6,454 & 59,712 \\
Ruby                                                                    & 29                       & 3,116 & 171   & 688   & 1,740 & 3,429 & 19,847 \\
\midrule
Total                                                                   & 158                      &       &       &       &       &       &       \\
\bottomrule
\end{tabular}
\end{center}
\end{table}

\subsubsection{Corpus for RQ3}

As previously discussed, RQ3 analyzes the number of defects and the time required to fix them before and after the adoption of feature toggles. Thus, analogously to the previous research question, the corpus for this question should only contain projects with enough commits (before and after) to allow at least one issue or pull request classified as a defect.

\begin{figure}
  \centering
  \caption{Distribution of commits per defect in projects with at least one defect in $C_{RQ1}$. The boxplot shows that all projects with at least 132 commits have at least one reported defect, except for outliers}.
  \label{fig:fig3}
\includegraphics[width=0.7\textwidth]{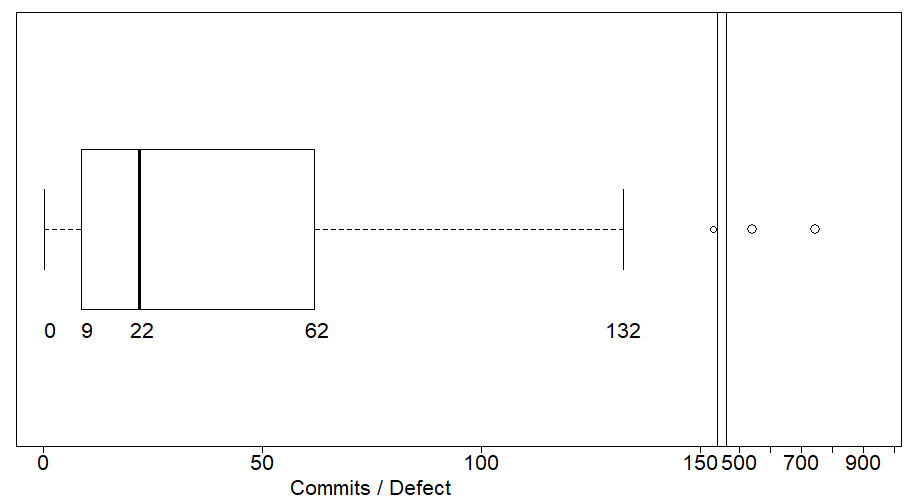}
\end{figure}

Aiming at finding such commit threshold, we first selected from $C_{RQ1}$ all projects that have at least one issue or pull request classified as a defect. This includes 163 projects in total. Then, we calculated the average number of commits per defect for each project. Figure~\ref{fig:fig3} shows the distribution of commits per defect. The upper limit of the boxplot is 132 commits per defect. The interpretation of such threshold is that all projects with at least 132 commits, except outliers, have at least 1 reported defect.

Finally, we applied this threshold to select projects with at least one reported defect that have at least 132 commits before and 132 commits after the adoption of feature toggles. This guarantees that any selected project has enough commits to be representative. As a result of this filter, we created a new corpus for RQ3, named $C_{RQ3}$, with 79 projects, as shown in Table~\ref{tbl:tbl5}.

\begin{table}
\caption{Distribution of the projects in $C_{RQ3}$}.
\label{tbl:tbl5}
\begin{center}
\begin{tabular}{lrrrrrrr}
\toprule
\multirow{2}{*}{\begin{tabular}[c]{@{}l@{}}Prog\\    Lang\end{tabular}} & \multirow{2}{*}{\# Proj} & \multicolumn{6}{c}{\# Commits}                 \\ \cline{3-8} 
                                                                        &                          & Mean  & Min   & Q1    & Med   & Q3    & Max    \\
\toprule
C\#                                                                     & 5                        & 2,846 & 415   & 1,910 & 3,214 & 3,499 & 5,196  \\
Java                                                                    & 14                       & 5,287 & 483   & 3,028 & 5,286 & 6,152 & 18,035 \\
JavaScript                                                              & 11                       & 1,997 & 523   & 675   & 1,379 & 1,986 & 7,001  \\
PHP                                                                     & 2                        & 2,054 & 2,046 & 1,562 & 2,054 & 2,062 & 2,071  \\
Python                                                                  & 31                       & 8,438 & 525   & 1,952 & 2,206 & 5,570 & 59,712 \\
Ruby                                                                    & 16                       & 4,084 & 507   & 1,270 & 1,985 & 3,990 & 19,847 \\
\midrule
Total                                                                   & 79                       &       &       &       &       &       &   \\
\bottomrule
\end{tabular}
\end{center}
\end{table}

\subsection{Control corpus}
We acknowledge that many of the data that we intend to analyze in this paper may be impacted by  factors other than the use of a feature toggle framework. For example, the number of merges may naturally grow over time as a project becomes more mature. Another example would be the time to fix a defect becoming smaller as a project attracts more contributors. Aiming at minimizing this bias, we build a separate corpus, which we call $C_{Control}$, to use as control during the analyses.

The universe of candidate projects to compose $C_{Control}$ was collected with the GitHub API in April/2021. It comprises 21,150 public non-fork and active \footnote{A repository was considered active if it contained at least one commit in the last 3 months from the collection date.} repositories, with at least 1,000 stars. The number of projects to be used was defined as around 10\% of the number of projects in each language in $C_{RQ1}$. Selected projects are inspected to ensure that they do not use any feature toggles frameworks. We tried to select relevant projects that combine the characteristics of both $C_{RQ2}$ and $C_{RQ3}$. Thus, considering that $C_{RQ2}$ was filtered based on the number of commits, selected control projects should have at least the minimum amount of commits for projects in $C_{CR2}$, in this case, 171. Furthermore, considering that $C_{RQ3}$ was filtered to guarantee the presence of issues, selected control projects should also have at least the minimum amount of issues in $C_{RQ3}$, in this case 1. In total, 98 projects that meet these criteria were randomly selected. One of these projects (nodejs/node-chakracore) was found to be too big to be analyzed under our time constraints. Consequently, the final number of repositories in $C_{Control}$ is 97. Its distribution of commits at the data collection time is displayed in Table \ref{tbl:control_distribution}. These projects have a representative distribution of commits, when compared to this study's corpora ($C_{RQ1}$, $C_{RQ2}$, and $C_{RQ3}$).

\begin{table}[]
\caption{Distribution of projects in $C_{Control}$.}
\label{tbl:control_distribution}
\begin{center}
\begin{tabular}{@{}lrrrrrrrr@{}}
\toprule
\multirow{2}{*}{\begin{tabular}[c]{@{}l@{}}Prog.\\    Lang\end{tabular}} & \multirow{2}{*}{\# Proj} & \multicolumn{7}{c}{\# Commits}         \\ \cline{3-9} 
                                                                        &                          & Mean & Std. Dev.  & Min & Q1 & Med & Q3   & Max    \\
\toprule
C\#        & 9       & 2,331       & 2,383   & 256 & 589   & 1,138  & 3,489  & 7,238  \\
Java       & 20      & 957      & 1,031   & 173 & 391 & 640 & 979 & 4,558  \\
JavaScript & 37      & 1,090       & 948       & 174 & 294   & 853   & 1,625  & 3547  \\
PHP        & 2       & 566        & 102       & 494 & 530   & 566   & 602   & 638   \\
Python     & 16      & 1,311       & 1,471      & 221 & 412   & 532   & 1,479  & 4,185  \\
Ruby       & 13      & 3,508       & 5,489      & 252 & 630   & 1,021  & 2,265  & 15,993 \\ \midrule
Total      & 97      &            &           &     &       &       &       &       \\ \bottomrule
\end{tabular}
\end{center}
\end{table}

The data collection for the control projects followed the same procedures described in the previous sections. The only difference was regarding the splitting of the project commit history. Since these projects do not use feature toggles frameworks, we used the middle of their commit history as a segregation point. Thus, for every project, we divided their history in two groups: first half of the history (first 50\% of the commits) and last half of the history (last 50\% of the commits). The first half of the history includes commits from the beginning of the project's history until the segregation point. The last half of the history includes commits from the segregation point until 31st May 2018, which is the date when the $C_{Cleaned}$ data was collected. Using the collected data for the two groups in each $C_{Control}$ project, we performed the same comparisons that were performed for $C_{RQ2}$ and $C_{RQ3}$, to check if the effects observed for those projects were impacted by the use of feature toggles frameworks or by the natural evolution of the projects. The results will be discussed in the next section.

\section{Results and Discussions}\label{sec:results}

In this section, we answer each research question. We also discuss the obtained results and the primary outcomes of our research.

\subsection{What is the adoption level of feature toggles in open-source projects (RQ1)?}

In this research question, we analyzed the popularity of each feature toggles framework and the moment of adoption of feature toggles frameworks on the projects.

Table~\ref{tbl:adoption} shows data regarding the number of projects using each of the frameworks, considering the whole life of each project and how many adopted each framework in the last year of our collected data.

\begin{table}
\caption{Feature toggles frameworks adoption.}
\label{tbl:adoption}
\begin{center}
\begin{tabular}{@{}llrrr|rr@{}}
\toprule
& & \multicolumn{3}{c}{Overall} & \multicolumn{2}{c}{Last year} \\ \midrule
Framework                 
& \multicolumn{1}{c}{\begin{tabular}[c]{@{}l@{}}Prog. \\ language \end{tabular}}

& \multicolumn{1}{c}{\begin{tabular}[c]{@{}l@{}} \# Projects\end{tabular}}
& \multicolumn{1}{c}{\begin{tabular}[c]{@{}l@{}} \% Adoption\end{tabular}}
& \multicolumn{1}{c}{\begin{tabular}[c]{@{}l@{}}\% Adoption  \\ within prog. \\ language\end{tabular}}
& \multicolumn{1}{c}{\begin{tabular}[c]{@{}l@{}} \# Projects \end{tabular}}
& \multicolumn{1}{c}{\begin{tabular}[c]{@{}l@{}} \% Adoption \end{tabular}} \\ \midrule
angular-toggle-switch     & JavaScript & 342 & 36.0\%  & 91.7\%  & 48  & 26.4\%  \\
Togglz                    & Java       & 153 & 16.1\%  & 77.7\%  & 48  & 26.4\%  \\
django-waffle             & Python     & 120 & 12.6\%  & 79.5\%  & 22  & 12.1\%  \\
featureflags              & Ruby       & 74  & 7.8\%   & 57.4\%  & 6   & 3.3\%   \\
FeatureToggle             & C\#        & 54  & 5.7\%   & 63.5\%  & 23  & 12.6\%  \\
Rollout                   & Ruby       & 53  & 5.6\%   & 41.1\%  & 5   & 2.7\%   \\
FF4J                      & Java       & 44  & 4.6\%   & 22.3\%  & 13  & 7.1\%   \\
Gargoyle                  & Python     & 17  & 1.8\%   & 11.3\%  & 1   & 0.5\%   \\
Switcheroo                & C\#        & 16  & 1.7\%   & 18.8\%  & 2   & 1.1\%   \\
FeatureSwitcher           & C\#        & 15  & 1.6\%   & 17.6\%  & 1   & 0.5\%   \\
ember-feature-flags       & JavaScript & 14  & 1.5\%   & 3.8\%   & 3   & 1.6\%   \\
Qandidate   Toggle        & PHP        & 14  & 1.5\%   & 100.0\% & 6   & 3.3\%   \\
Flask-FeatureFlags        & Python     & 11  & 1.2\%   & 7.3\%   & 2   & 1.1\%   \\
Ericelliot/feature-toggle & JavaScript & 10  & 1.1\%   & 2.7\%   & 2   & 1.1\%   \\
fflip                     & JavaScript & 7   & 0.7\%   & 1.9\%   & 0   & 0.0\%   \\
Gutter                    & Python     & 3   & 0.3\%   & 2.0\%   & 0   & 0.0\%   \\
feature\_flipper          & Ruby       & 2   & 0.2\%   & 1.6\%   & 0   & 0.0\%   \\ \midrule
Total                     &            & 949 & 100.0\% &         & 182 & 100.0\% \\ \bottomrule
\end{tabular}
\end{center}
\end{table}

Table~\ref{tbl:adoption} shows that, when considering the whole lifetime of the projects, the most popular feature toggles frameworks are \textit{angular-toggle-switch}, \textit{Togglz}, and \textit{django-waffle}, being adopted by 342, 153, and 120 projects, respectively. They are also the most popular frameworks within their respective programming languages -- JavaScript, Java, and Python, respectively. Regarding Ruby programming language, there is no clear preferred framework. The \textit{featureflags} framework has been adopted by 57.4\% of the Ruby projects, while \textit{Rollout} has been adopted by 41.1\%.

Considering only the last year of the projects' data (last two columns in Table \ref{tbl:adoption}), note that despite \textit{angular-toggle-switch} is 2.23 times more adopted than \textit{Togglz} overall, their adoption in the last year was the same, both with 26.4\% of the total. If we compare the overall percentage of adoption with last year's adoption, we can observe that the popularity of \textit{angular-toggle-switch} decreases (from 36\% to 26.4\%), while for \textit{Togglz} it increases (from 16.1\% to 26.4\%). For the \textit{django-waffle} framework, the popularity did not change significantly (from 12.6\% to 12.1\%).

\begin{figure}
    \centering
    \includegraphics[width=0.55\textwidth]{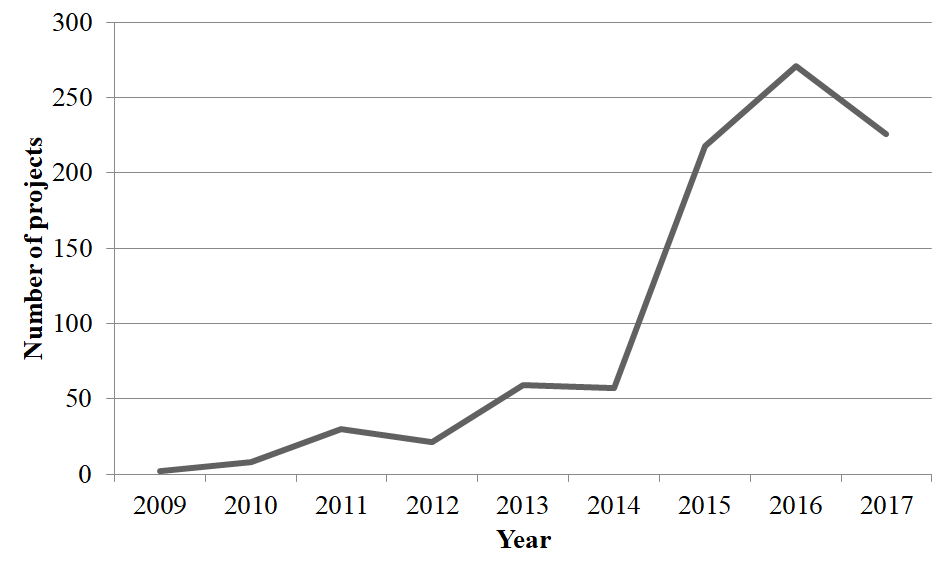}
    \caption{Adoption of feature toggles frameworks over the years for projects in $C_{RQ1}$}.
    \label{fig:adoption}
\end{figure}

Figure~\ref{fig:adoption} shows how the adoption of feature toggles frameworks has changed over the years\footnote{The year of 2018 was excluded since our data collection was in May/2018.}. It is possible to observe that 2015 had a very high adoption compared to the previous years, with the peak being reached in 2016. This high adoption can be explained by the release of the most popular framework of $C_{RQ1}$, \textit{angular-toggle-switch}, in February 2015. In that year alone, 106 projects adopted it, while in 2016, the adoption increased to 145 new projects.

As we mentioned in Section~\ref{sec:materials}, we identified the exact moment when feature toggles were adopted in each project. Thus, we could observe that 667 projects (70\% of $C_{RQ1}$) were created without feature toggles and, at some point, introduced a feature toggles framework. The average milestone of adopting a feature toggles framework was 982 commits after the project creation, with a standard deviation of 4,771 commits. This average adoption milestone is located on average at 44\% of the commits history of the projects contained in our collected snapshot. It is important to notice that this percentage will vary if more commits are included in the corpus in the future. However, it should give an overall idea of the point where we divide the history of the projects for the analyses in RQ2 and RQ3.

When contrasting the projects that used feature toggles since their first commit (30\% of $C_{RQ1}$) and the projects that adopted feature toggles afterward (70\% of $C_{RQ1}$), we can observe that the average number of developers and the average number of commits differ, as detailed in Table~\ref{tbl:tbl7}. This clearly shows that the bigger and more mature projects are the ones that adopted feature toggles after their creation. This is natural, considering that feature toggle is a recent technology. 

\begin{table}
\caption{Comparison of projects that always used feature toggles and projects that adopted it afterwards.}
\label{tbl:tbl7}
\begin{center}
\begin{tabular}{lrrrrr}
\toprule 
\begin{tabular}[c]{@{}l@{}}Adoption \\    Moment\end{tabular} & \# Projects & \begin{tabular}[c]{@{}l@{}}Mean\\    \# Developers\end{tabular} & \begin{tabular}[c]{@{}l@{}}Mean \\    \# Commits\end{tabular} & \begin{tabular}[c]{@{}l@{}}Mean\\    Normalized \\    \# Merges*\end{tabular} & \begin{tabular}[c]{@{}l@{}}Mean \\    Normalized \\    \# Defects*\end{tabular} \\
\toprule
Beginning                                                     & 282         & 2                                                               & 32                                                            & 8.13                                                                          & 10.29                                                                           \\
Afterwards                                                    & 667         & 22                                                              & 1,438                                                         & 9.40                                                                          & 2.08                                                                            \\
\bottomrule 
\multicolumn{6}{l}{\begin{tabular}[c]{@{}l@{}}* Considering only projects with registered merges (beginning: 43; afterwards: 371) \\ and  defects (beginning: 18; afterwards: 145).\end{tabular}}        
\end{tabular}
\end{center}
\end{table}

We can also observe in Table~\ref{tbl:tbl7} that projects using feature toggles since the beginning have, on average, fewer merges and more defects. We proceeded with hypothesis tests over the number of developers, number of commits, normalized number of merges, and normalized number of defects. For each variable, we divide the data into two groups, one containing data from projects that adopted feature toggles since their creation and another with data from the remaining projects. A summary of the obtained measures during our hypothesis testing is displayed in Table~\ref{tbl:rq1_hypothesis_summary}. We first run a Shapiro-Wilk's test to check whether the data for each variable in each group follows a normal distribution, considering $\alpha = 0.05$. According to the values displayed in the Shapiro Wilk's p-value column in Table~\ref{tbl:rq1_hypothesis_summary}, none of the analyzed variables' data follow a normal distribution. Thus, we applied the Mann-Whitney test \cite{mann_test_1947_22}, an unpaired nonparametric test for two independent samples, and observed a statistically significant difference ($p-value \leq 0.05$) for the number of developers, the number of commits, and the normalized number of defects. We could not observe a statistically significant difference for the normalized number of merges.

\begin{table}[]
\begin{center}
\caption{Summary of measures obtained during hypothesis tests comparing variables from projects that adopted feature toggles since their creation and projects that adopted it afterwards.}
\label{tbl:rq1_hypothesis_summary}
\begin{tabular}{@{}lrrrr@{}}
\toprule
\multicolumn{1}{c}{\multirow{2}{*}{\textbf{Variable}}} & \multicolumn{2}{c}{\textbf{Shapiro-Wilk's p-value}}                              & \multicolumn{1}{c}{\multirow{2}{*}{\begin{tabular}[c]{@{}c@{}}\textbf{Mann-Whitney's} \\ \textbf{p-value} \end{tabular}}} & \multicolumn{1}{c}{\multirow{2}{*}{\textbf{Cliff's Delta (d)}}} \\
\multicolumn{1}{c}{}                                   & \multicolumn{1}{c}{\textbf{Beginning}} & \multicolumn{1}{c}{\textbf{Afterwards}} & \multicolumn{1}{c}{}                                                 & \multicolumn{1}{c}{}                                            \\ \midrule
\# of developers                                       & $3.388 \times 10^{-16}$                & $3.388 \times 10^{-16}$                 & $2.2 \times 10^{-16}$                                                & $0.487$                                                         \\
\# of commits                                          & $3.388 \times 10^{-16}$                & $3.388 \times 10^{-16}$                 & $2.2 \times 10^{-16}$                                                & $0.671$                                                         \\
normalized \# of merges                                & $0.023$                                & $3.388 \times 10^{-16}$                 & $0.684$                                                              & $0.038$                                                         \\
normalized \# of defects                                & $4.723 \times 10^{-4}$                 & $3.388 \times 10^{-16}$                 & $8.838 \times 10^{-3}$                                               & $0.379$                                                         \\ \bottomrule
\end{tabular}
\end{center}
\end{table}

Then, we decided to verify the magnitude of the difference for each variable between the groups of projects that adopted feature toggles since their creation and projects that adopted it afterward. To assess this difference, we calculate the effect size \cite{sullivan_using_2012_27}. As the samples did not follow a normal distribution, we applied Cliff’s Delta, a nonparametric effect size method for two samples \cite{macbeth_cliffs_2011_23}. We used Romano’s thresholds \cite{romano_exploring_2006_24} to interpret the effect size of d: $|d| < 0.147$ indicates “negligible effect”, $0.147\leq |d| < 0.330$ indicates “small effect", $0.330\leq |d| < 0.474$ indicates “medium effect”, and $0.474 \leq |d|$ indicates “large effect”. Thus, according to the values displayed in the Cliff's Delta (d) column in Table~\ref{tbl:rq1_hypothesis_summary}, we observed a large effect size for the number of developers and the number of commits, a medium effect size for the normalized number of defects, and a negligible effect size for the normalized number of merges. Based on this, we can conclude that projects that used a feature toggles framework since their creation are much smaller in terms of the number of developers and commits in comparison to projects that adopted it afterward. In addition, they have more defects. Despite having fewer merges, we could not confirm that the effect of using feature toggles since their creation had any impact on this result, since the difference was not significant.

In addition to the moment of adoption, we also checked if the projects continued to use a feature toggles framework after adopting it. For this analysis, we counted the number of references in the source code to each feature toggles framework on each project. If the number of references in the current version is greater than in the adoption version, then we say that the usage increased. If it is lesser, we say that the usage decreased. If equal, we say that the usage stayed the same, and if there is no reference in the current version, we say that the project stopped using it. In total, we performed this analysis on 849 of the 949 projects from $C_{RQ1}$. Not every project was used due to missing or changed history of these projects, since this analysis was made some time after the data collection. Overall, we could observe that 75.9\% of the projects did not change the usage of the frameworks. In 19.8\% of the projects, the usage increased, whereas in 4.2\%, it decreased. Only one project (0.1\%) stopped using the detected framework. These numbers evidence that once adopted, most projects continue to use the detected framework throughout its history.

\begin{center}
\noindent\fbox{%
    \parbox{0.985\textwidth}{%
        \textbf{Finding 1}: We could find just 949 projects in GitHub using a feature toggles framework. This number is small considering the total number of projects in GitHub. Most of these projects (70\%) were created without a feature toggles framework and adopted this technology on average, after 982 commits of their creation. They also tend to continue using a framework once it is adopted. The projects that used a feature toggles framework since their creation (30\%) are small in terms of the number of developers and commits, and present fewer merges and more defects. However, we could not confirm that having fewer merges had any relationship with the use of feature toggles since their beginning. Overall, JavaScript is the most popular language (26\%) among projects that use a feature toggles framework, followed by Java (15\%) and Python (12\%).
    }%
}
\end{center}

\subsection{Do the number of branch merges and the necessary effort change after adopting feature toggles (RQ2)?}

In this research question, we aim at checking if the number or the effort of merges changes significantly after the adoption of feature toggles. Table \ref{tbl:rq2_summary} displays a summary of the obtained measures for the sub questions in RQ2. The measures obtained for $C_{Control}$ are also displayed for comparison. The results for each sub question is discussed in the next sections.

\begin{table}[]
\begin{center}
\caption{Summary of statistics for sub questions of RQ2 for projects in $C_{RQ2}$ before and after adopting FT, and before and after the segregation point (SP) for projects in $C_{Control}$.}
\label{tbl:rq2_summary}
\begin{tabular}{lllrrrr}
\toprule
\multirow{2}{*}{RQ}  & \multicolumn{1}{c}{\multirow{2}{*}{Measure}}            & \multirow{2}{*}{}           & \multicolumn{2}{c}{$C_{RQ2}$}               & \multicolumn{2}{c}{$C_{Control}$} \\ \cmidrule{4-7} 
                     &                                     &                             & Before FT & After FT                   & Before SP        & After SP       \\ \midrule
\multirow{2}{*}{2.1} & \multirow{2}{*}{Number of merges}   & \multicolumn{1}{l|}{Mean}   & 10.72     & \multicolumn{1}{r|}{10.16} & 2.94          & 4.15        \\
                     &                                     & \multicolumn{1}{l|}{Median} & 9.40       & \multicolumn{1}{r|}{8.10}   & 2.19          & 2.27        \\
\multirow{2}{*}{2.2} & \multirow{2}{*}{Effort per merge}   & \multicolumn{1}{l|}{Mean}   & 13.34     & \multicolumn{1}{r|}{1.68}  & 22.15         & 3.93        \\
                     &                                     & \multicolumn{1}{l|}{Median} & 0.56      & \multicolumn{1}{r|}{0.04}  & 0.01          & 0.10         \\
\multirow{2}{*}{2.3} & \multirow{2}{*}{Total merge effort} & \multicolumn{1}{l|}{Mean}   & 87.58     & \multicolumn{1}{r|}{17.97} & 195.37          & 26.27        \\
                     &                                     & \multicolumn{1}{l|}{Median} & 3.86      & \multicolumn{1}{r|}{0.40}   & 0.04         & 0.26        \\ \bottomrule
\end{tabular}
\end{center}
\end{table}

\subsubsection{Number of merges (RQ2.1)}\label{sec:number_merges}

\begin{figure}
  \centering
  \caption{Distribution of the normalized number of merges, before and after adopting feature toggles (FT). The boxplot shows that the median number of merges before the adoption of FT is 9.4. After the adoption of FT it decreased to 8.1.}
  \label{fig:fig4}
\includegraphics[width=0.7\textwidth]{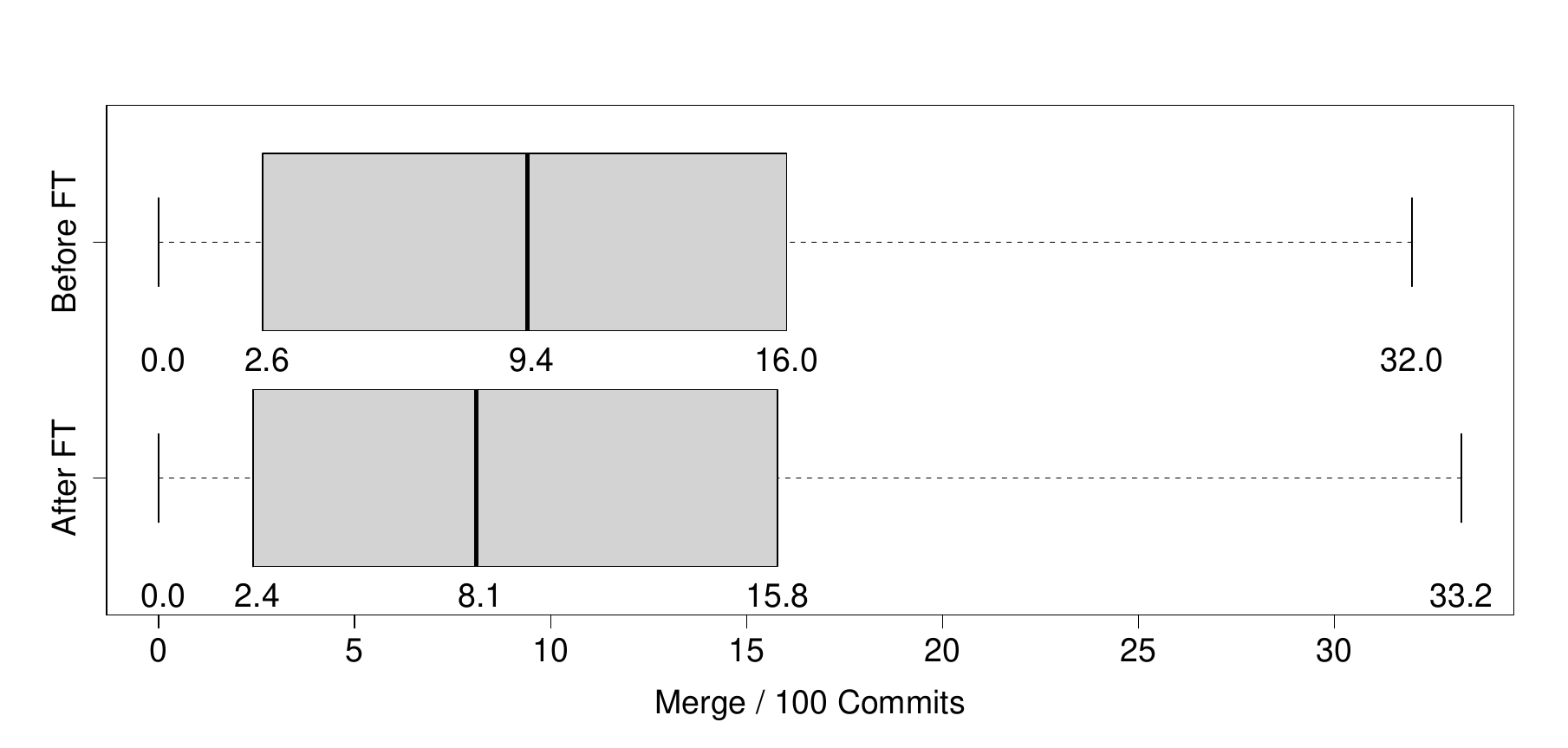}
\end{figure}

As previously mentioned, the following analysis is based on $C_{RQ2}$, which is the corpus of repositories with enough commits (i.e., 82) for having at least one merge before and one merge after the adoption of feature toggles. In this analysis, we focus on the normalized number of merges of the projects in $C_{RQ2}$ across two different periods for each project, before and after adopting a feature toggles framework. Figure~\ref{fig:fig4} displays boxplots with the normalized number of merges before and after adopting feature toggles.

\begin{figure}
  \centering
  \caption{Scatter plot with the number of merges per 100 commits, before and after adopting feature toggles (FT).}
  \label{fig:fig5}
\includegraphics[width=0.7\textwidth]{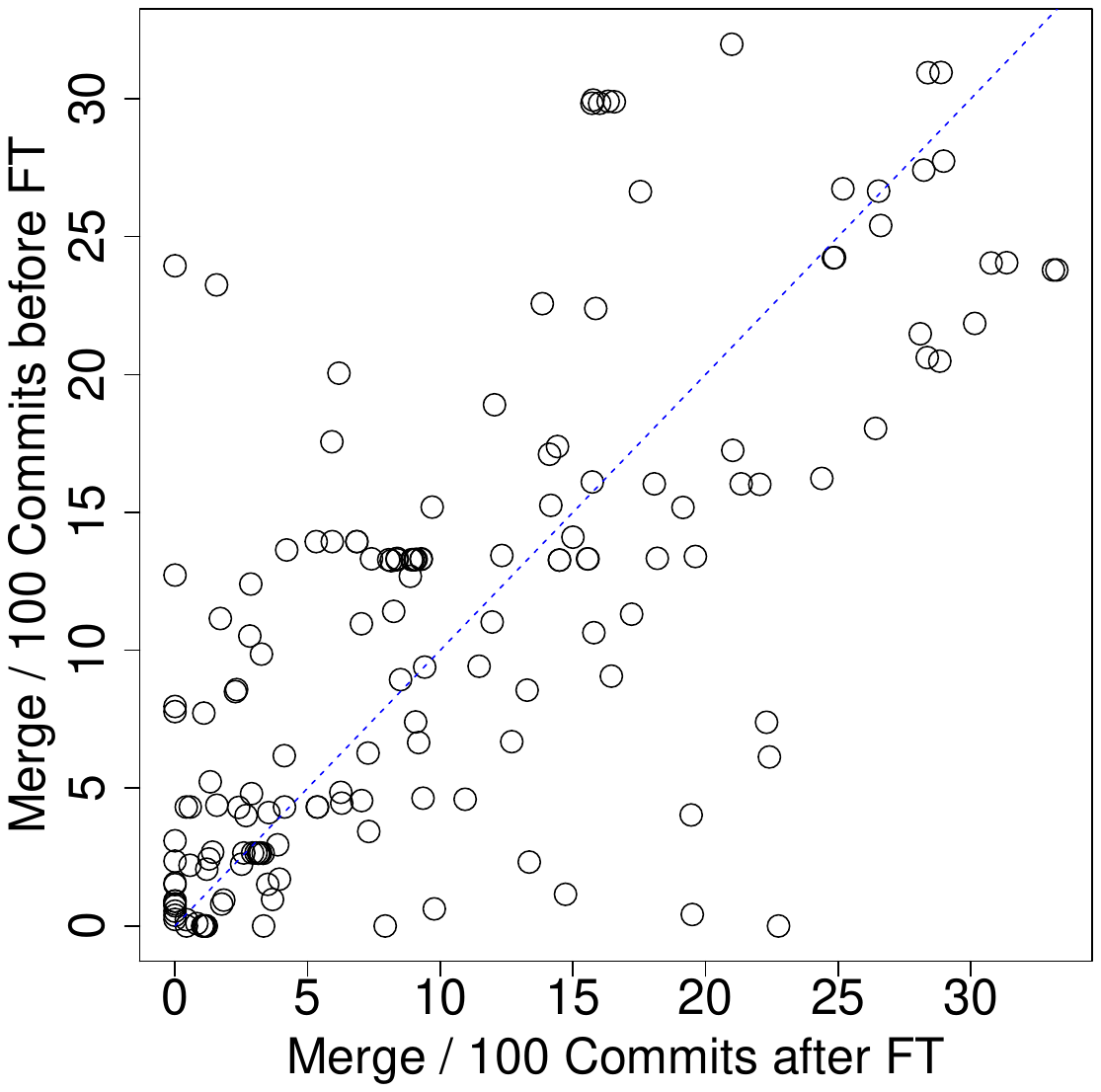}
\end{figure}

A visual inspection of the boxplots in Figure~\ref{fig:fig4} shows very similar distributions for the two groups of the analyzed data. However, we run a hypothesis test to check if there is any  statistically significant difference between the number of merges before and after adopting FT. We first run a Shapiro test to check whether the data for both groups follows a normal distribution, considering $\alpha = 0.05$. We found that they do not follow a normal distribution, with $p\mbox{-}value = 5.225 \times 10^{-08}$ for commits before feature toggle and $p\mbox{-}value = 3.732 \times 10^{-09}$ for commits after feature toggles. Thus, we applied the Wilcoxon paired test \cite{wilcoxon_individual_1945_25} and observed $p\mbox{-}value = 0.3292$, which indicates that there is no significant difference between the samples. Figure~\ref{fig:fig5} shows a scatter plot based on the number of merges per 100 commits, before and after introducing feature toggles. Most of the projects are concentrated in the bottom-left quadrant, which indicates few merges per 100 commits. Furthermore, in general, they are very concentrated near the diagonal of the chart, which shows a linear correlation among the samples. This linear correlation suggests that the number of merges before and after the adoption of feature toggles seems to be equivalent.

Therefore, despite the possibility of isolating features without feature branches, we could not observe a significant reduction in the number of merges after adopting feature toggles. Table~\ref{tbl:rq2_summary} shows the mean and median number of merges per 100 commits.

\begin{figure}
  \centering
  \caption{Comparison of the number of merges per 100 commits for each programming language.}
  \label{fig:fig6}
\includegraphics[width=1\textwidth]{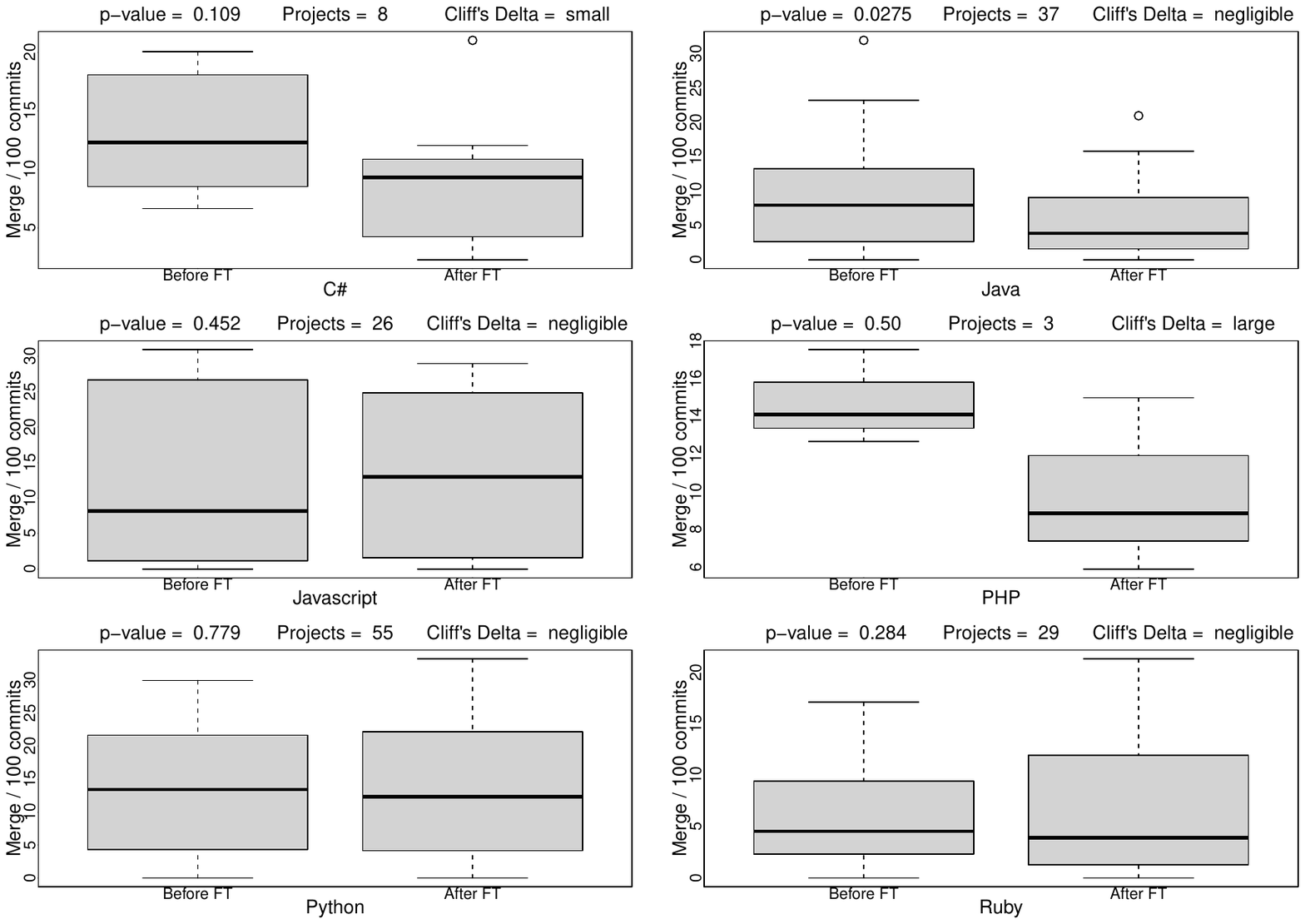}
\end{figure}

Finally, we also checked whether the result is the same for each programming language. Figure~\ref{fig:fig6} displays the distribution for each programming language and the respective p-value. None of the programming languages presented a normal distribution and, consequently, we applied the Wilcoxon paired test \cite{wilcoxon_individual_1945_25} in all cases.

We can observe that there is no universal tendency among all programming languages. For C\#, Java, and PHP, the median number of merges reduces after adopting feature toggles. For JavaScript, we could observe the opposite situation. We could reject $H_0$ just for Java. However, such multiple comparisons can increase Type-I errors. To avoid this threat, if we apply Bonferroni correction\footnote{Bonferroni correction is an adjustment made to the alpha-value to mitigate the multiple comparison problem, thus reducing the chances of Type-I errors (i.e., false-positives). It consists of dividing the alpha-value by the number of comparisons.} \cite{bonferroni_teoria_1936_26}, our $\alpha$-value drops to $0.008 (0.05 \div 6)$ and $H_0$ is not rejected for Java anymore. For Python and Ruby, the average number of merges showed few changes.

To complement the results for this research question, we analyzed whether the number of merges changed after the segregation point for the projects in $C_{Control}$. The Shapiro-Wilk's normality test suggests that the distribution of both groups is not normal, with $p\mbox{-}value = 5.278 \times 10^{-08}$ for commits in the first half of the history and $p\mbox{-}value = 2.316 \times 10^{-12}$ for commits in the last half of the history. Thus, we employed the Wilcoxon paired test and obtained a $p\mbox{-}value = 0.025$, which indicates that there is a significant difference between the two groups regarding the number of merges for the control projects. Finally, we calculated the Cliff's Delta to check the magnitude of the differences and found a delta estimate of $d=-0.089$, which suggests that the observed increase in the number of merges is negligible.

Summarizing the results, we found no significant difference in the number of merges after the adoption of feature toggles in $C_{RQ2}$. On the other hand, there was a statistically significant increase in the number of merges for the control projects ($C_{Control}$), but with a negligible effect size. Consequently, we cannot say that the use of feature toggles frameworks has any impact on the number of merges.

\begin{center}
\noindent\fbox{%
    \parbox{0.985\textwidth}{%
        \textbf{Finding 2}: Although feature toggles enable trunk-based development, we could not observe significant changes in the number of merges, even when analyzing different programming languages. In contrast, we found a significant but negligible increase in the number of merges for the control projects. Consequently, we could not confirm that the use of feature toggles frameworks impacted on the number of merges.
    }%
}
\end{center}

\subsubsection{Effort per merge (RQ2.2)}\label{sec:effort_merge}

\begin{figure}
  \centering
  \caption{Distribution of effort per merge in $C_{RQ2}$, before and after adopting feature toggles. Some outliers were omitted to ease visualization. The boxplot shows that the median of the average merge effort was 0.56 before the adoption of FT. After the adoption of FT it decreased to 0.04.}
  \label{fig:fig7}
\includegraphics[width=0.7\textwidth]{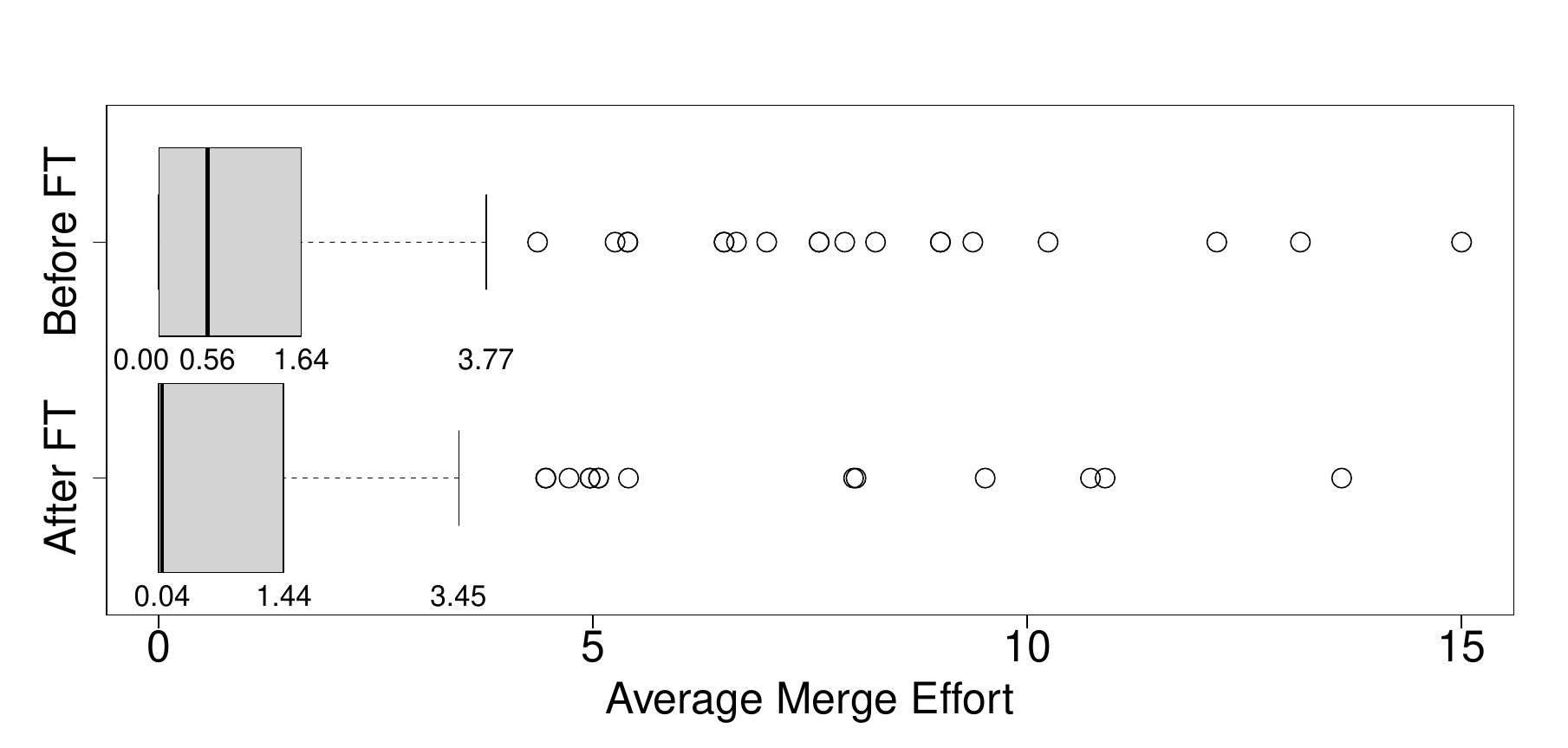}
\end{figure}

In this research question, we study whether the adoption of feature toggles changes the effort per merge. Thus, using $C_{RQ2}$, we calculated the effort for each merge. Then, we divided the calculated data into two groups, one containing the merge effort for merges that happened before the adoption of feature toggles, and the other for merges that happened after the adoption of feature toggles. The effort was measured in terms of added and removed lines of code, as explained in Section~\ref{sec:corpus}. Figure~\ref{fig:fig7} displays a boxplot with the distribution of effort per merge before and after the introduction of feature toggles. While, on average, each merge demanded 13.3 extra actions (lines added or removed) before adopting feature toggles, this number dropped to 1.7 afterward, as shown in Table~\ref{tbl:rq2_summary}. 

Proceeding similarly to the previous analysis (Section~\ref{sec:number_merges}), we first observed non-normality in the effort per merge data for both groups using the Shapiro test ($p\mbox{-}value < 0.001$). Then, we employed the Wilcoxon paired test to check if there is any significant difference between the data for the two groups. We observed a $p\mbox{-}value = 0.039$, indicating a significant difference. The reduction in the mean and median shown in Table~\ref{tbl:rq2_summary} are around 87\% and 93\%, respectively. Consequently, we checked the magnitude of the difference. As the data for the two groups do not follow a normal distribution, we applied Cliff’s Delta and observed a small effect size ($0.1889$). The interpretation of this result is that the observed decrease in the effort per merge before and after adopting FT is significant, although the magnitude of this difference is small.

\begin{figure}
    \centering
    \caption{Comparison of the average effort per merge for each programming language. Outliers are omitted for better visualization.}
    \includegraphics[width=1\textwidth]{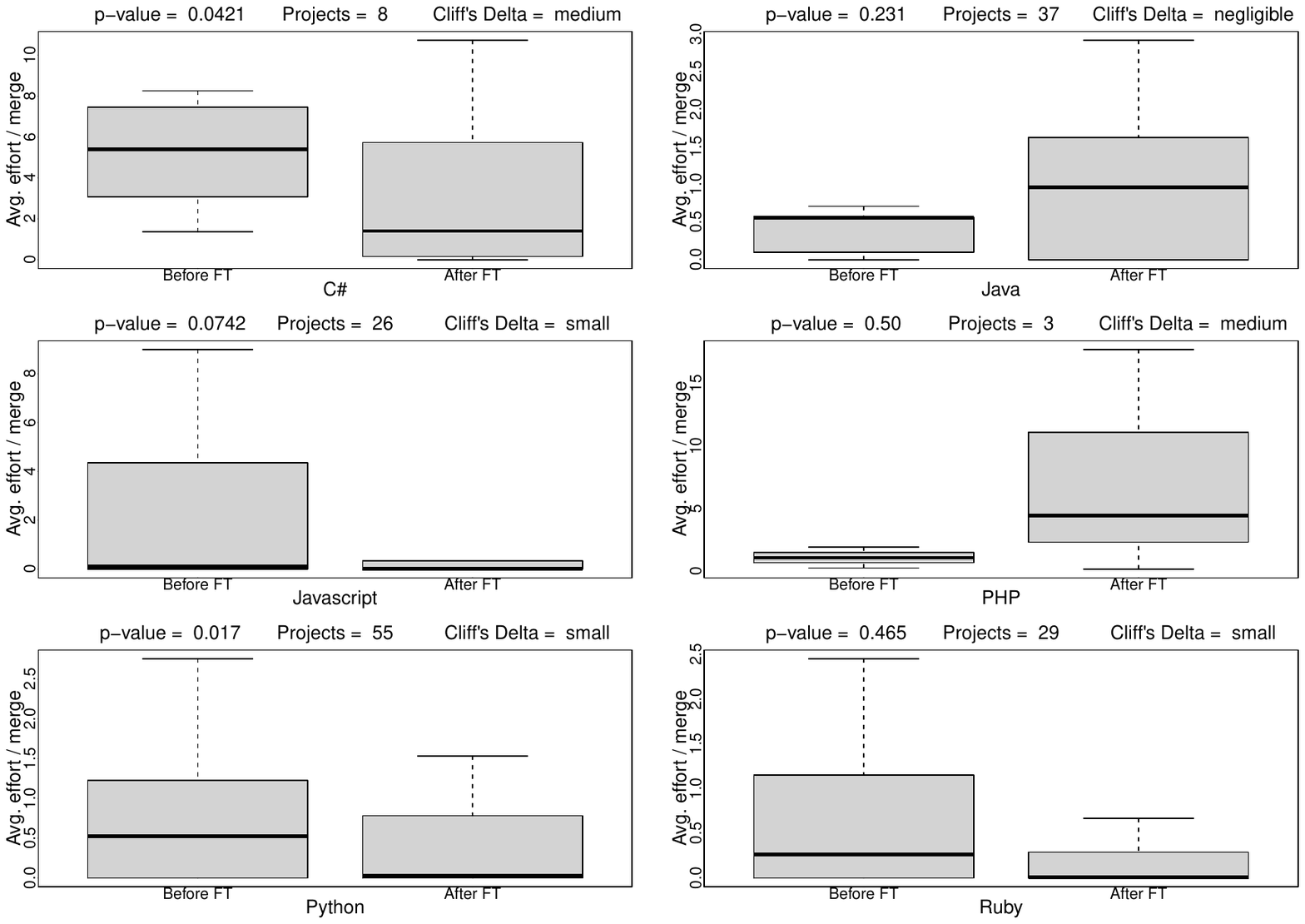}
    \label{fig:boxplot-prog-lang-rq22}
\end{figure}

As in Section~\ref{sec:number_merges}, we also segmented the analysis by programming language to investigate whether the results are uniform. Figure~\ref{fig:boxplot-prog-lang-rq22} shows the boxplots for each of the programming languages. According to the applied Shapiro-Wilk's test, none of the programming language samples presented a normal distribution, except for PHP. However, according to Warner \cite{warner2012applied}, nonparametric tests should be used for small samples, even without performing the normality and homoscedasticity tests. Thus, we applied the nonparametric Wilcoxon paired test \cite{wilcoxon_individual_1945_25} in all cases.

The average effort decreased for all languages, except for Java and PHP. For PHP this may be explained by the fact that we have only 3 projects. Despite this, considering the Bonferroni corrected $\alpha$-value of 0.008, the $H_0$ hypothesis is not rejected for any of the programming language samples. Thus, the p-values suggest that the differences in the samples are not statistically significant.

Aiming at further understanding the results, we qualitatively investigated the relevant cases with average merge effort variation when comparing the periods before and after the adoption of feature toggles frameworks.

To give an overall idea of how the average merge effort varied across the projects, in 51\% (80/158) of the cases it decreased after the adoption of a FT framework. Among these, 80\% (34/80) went from an average merge effort greater than zero to an average merge effort of zero. In 30\% (48/158) of the projects, the average merge effort has increased, while it stayed the same in 19\% (30/158) of them.

In two Mozilla Python projects (\textit{mozilla/zamboni} and \textit{mozilla/addons-server}), the average merge effort has increased by more than 100,000\%. This big increase, however, is due to the average merge effort before the adoption of FT being very low, with around 0.004 in both projects, increasing to 5.41 and 4.46, respectively, after the adoption of FT. It can be noted that both projects have a similar history before the adoption of FT. They both have 5,027 commits without FT and around 230 branch merges. This did change later, with the \textit{mozilla/addons-server} project achieving a bigger activity in both commits and merges than \textit{mozilla/zamboni}. Despite this, the high difference between the before and after FT average merge effort might be explained by the fact that in both projects around 90\% of the merges happened before the adoption of FT.

Ten more projects also showed an increase in the average merge effort by more than 1,000\%. By further inspection, we found that eight of them are unofficial forks of the Java project \textit{Frederikam/FredBoat}\footnote{ This repository has migrated to closed source after the data collection. The old URL is \url{https://github.com/Frederikam/FredBoat}. Despite this, its history can be inspected on unofficial forks, like \url{https://github.com/Siro256/FredBoat}.}. Since they are not official forks according to the GitHub API, they were not identified as forks during the corpus filtering. Thus, some of the merges of this project are replicated in our dataset. This might be the cause why Java showed an increase in the average merge effort, differently from the other languages. This project is a bot that can be used to play music in a voice chat application. Before adopting the \textit{Togglz} feature toggles framework, the average effort for its branch merges was 0.1. This number increased to 2.16 after the adoption of the framework. Analyzing each merge separately, we observed that, before the adoption, 29 branch merges occurred, most of them (27) with zero merge effort. On the other hand, after the adoption, 13 branch merges happened, whereas 12 had zero merge effort and one\footnote{\url{https://github.com/Frederikam/FredBoat/commit/6f73c3fc53bbc1b3e5b1a3b62f2eb1b18719d8a8}} had an effort value of 26. Further inspecting this merge reveals that it results from the merge of a branch that had not been synchronized for 27 days. Consequently, many conflicts had to be manually resolved. In this project, the average effort may not be proportional because of this big merge. The other two projects are \textit{kilimchoi/teamleada.com} and \textit{HabitantsLieuxMemoires/web-app}. Both are Ruby web-app projects with educational purposes. \textit{kilimchoi/teamleada.com} has an average merge effort of 0.84 before and 29.24 after the adoption of FT. The 3,397\% increase can be explained by one big merge\footnote{\url{https://github.com/kilimchoi/teamleada.com/commit/087c405bfb7ce3477c29343ce70594686ccd7692}} with an effort of 5,308 which happened in the period after the adoption of FT. In comparison, the highest merge effort before the adoption of FT is 68\footnote{\url{https://github.com/kilimchoi/teamleada.com/commit/7dd86d82feb09d0cf0034c5aa3d8ea3f92031236}}. On the other hand, \textit{HabitantsLieuxMemoires/web-app} has an average merge effort of 0.05 before and 0.96 after the adoption of FT. Thus, the 1,732\% increase can be explained by the very low merge effort in the before FT period.

To contrast with our analysis of projects where the average merge effort greatly increased, we also analyzed projects where it decreased. A total of 30 projects had a 100\% decrease. Most of them have zero average merge effort after the adoption of FT because there are none or just a few merges in this period. However, for this qualitative analysis, we will first inspect projects with zero average merge effort after the adoption of FT that have a balanced number of merges before and after the adoption of FT.


\textit{mozilla-services/socorro} is a Python project that provides tools for analyzing crash reports. It contains 904 merges before the adoption of FT, with an average merge effort of 0.3, and 941 merges after, with an average merge effort of zero. Overall, this project has only 6 merges with merge effort greater than zero, all of which happened before the adoption of FT. In these merges, the average effort is 4.83, with the highest being 14 and the lowest 1. All other 1,839 merges have 0 merge effort. \textit{CenterForOpenScience/isp} is a JavaScript web-app project aiming to support an experimental study. It has 109 merges before the adoption of FT with an average merge effort of 0.14, and 135 merges after, with an average merge effort of zero. It has only 4 merges with effort greater than zero, all of which happened before the adoption of FT. The average merge effort for these merges is 4, with the lowest being 2 and the highest 7. The last analyzed project in this batch is \textit{mozilla/mozillians}, which is also a Python web-app, but to connect Mozilla contributors. It has 390 merges before the adoption of FT, with an average merge effort of 0.53, and 373 merges after, with an average merge effort of zero. It has only 10 of its 763 merges with a merge effort greater than zero. Among these, the average merge effort is 20.8, with the lowest being four merges with effort 1 and the highest 111\footnote{\url{https://github.com/mozilla/mozillians/commit/2ff6986544ffbf6645935345a87e6acd514b17cc}}. These three projects follow the same pattern, with a very little proportion of merges that required any effort, all of which happened before the adoption of FT.

There is another spectrum of projects, where the average merge effort has decreased, but not to zero. Similar to the previous projects, for this qualitative analysis we also picked those that have a balanced number of merges before and after the adoption of FT. Unfortunately, the balance on this set of projects is smaller, so we picked projects which had a substantial merge effort before the adoption of FT and where the proportion of merges is at least 50\%.

\textit{contactbiin/BiinBackend} is a JavaScript Content Management System (CMS) project. In total, it has 342 merges before the adoption of FT and 268 after. Overall, in 25 merges the effort is greater than zero. From these, 24 happened before the adoption of FT, with an average effort of 307, and 1 happened after, with an effort of 3. From the merges that happened before the adoption, 7 of them have an effort greater than 100, with the highest one being 5,540\footnote{\url{https://github.com/contactbiin/BiinBackend/commit/4928ab244d1adac1d9e670993887c8d3a235fb0b}}. This behavior was similar in the \textit{SkillsFundingAgency/das-apprenticeship-programs-api} C\# project. It provides an API to search for apprenticeship programs. This project has 39 merges in total. 23 merges happened before the adoption of FT, with an average merge effort of 2.39. From these 23 merges, 4 had effort greater than zero, with an average of 13. For the period after the adoption of FT, 16 merges were performed, with an average effort of 0.12. From these, only one merge has effort greater than zero, with a value of 2. In these two projects, we observed a pattern where only a few merges have effort greater than zero, and most of them happened before the adoption of FT.

In contrast, in the JavaScript project \textit{madnight/gitter} which is a chat and network platform, many (798) merges with effort greater than zero happened. However, despite this, it also has a lot of merges overall (10,816). The number of merges with effort greater than zero is balanced between the before and after the adoption of FT, with 383 merges (with average effort of 128) before and 415 (with average effort of 46) after. Finally, a similar behavior was also observed in the \textit{ministryofjustice/manchester\_traffic\_offences\_pleas} Python project. It is a web-app for appealing traffic offences. It has a total of 330 merges, of which 40 have effort greater than zero, where 27 of them happened before the adoption of FT, with an average effort of 39.8, and 13 merges happened after the adoption of FT, with an average effort of 14.8. Overall, considering all merges, the merge effort decreased from 7.60 to 4.97.

From this qualitative analysis, we could observe that merges with zero merge effort are very common, at least for the cases we analyzed. Despite this, we could observe a bigger prevalence of these merges in the period after the adoption of FT. This suggests that the adoption of feature toggles frameworks might have influenced the overall average merge effort decrease. To check this, we compare our results with the results obtained in the control projects.

We analyzed if the effort for each merge has changed before and after the segregation point for the projects in $C_{Control}$. The Shapiro-Wilk's normality tests resulted in $p\mbox{-}value < 2.200 \times 10^{-16}$ for the merge effort data of both groups. Thus, we employed the nonparametric Wilcoxon paired test to check if there is a significant difference between both groups, resulting in a $p\mbox{-}value = 0.683$. The obtained results indicate that there is no significant difference between the data of average effort before and after the segregation point for projects in $C_{Control}$.

To summarize, we could observe a significant difference in the average merge effort for both groups in $C_{RQ2}$ but not in $C_{Control}$. This result suggests that the adoption of feature toggles frameworks might have influenced the average effort decrease observed in $C_{RQ2}$.

\begin{center}
\noindent\fbox{
    \parbox{0.985\textwidth}{
        \textbf{Finding 3}: We could observe a statistically significant difference in the effort for each merge before and after the adoption of feature toggles frameworks. This difference was not significant in the control projects. Both the mean and the median have decreased after adopting feature toggles, with a small effect size. By inspecting relevant cases, we found a bigger prevalence of merges with zero or close to zero effort in the period after the adoption of feature toggles frameworks.
    }
}
\end{center}

\subsubsection{Total merge effort (RQ2.3)}\label{sec:total_merge_effort}

\begin{figure}
  \centering
  \caption{Distribution of the normalized merge effort in $C_{RQ2}$, before and after adopting feature toggles. Some outliers were omitted for better visualization. The boxplot shows that the median of the normalized merge effort was 3.86 before the adoption of FT. After the adoption of FT, the median decreased to 0.40.}
  \label{fig:fig8}
\includegraphics[width=0.7\textwidth]{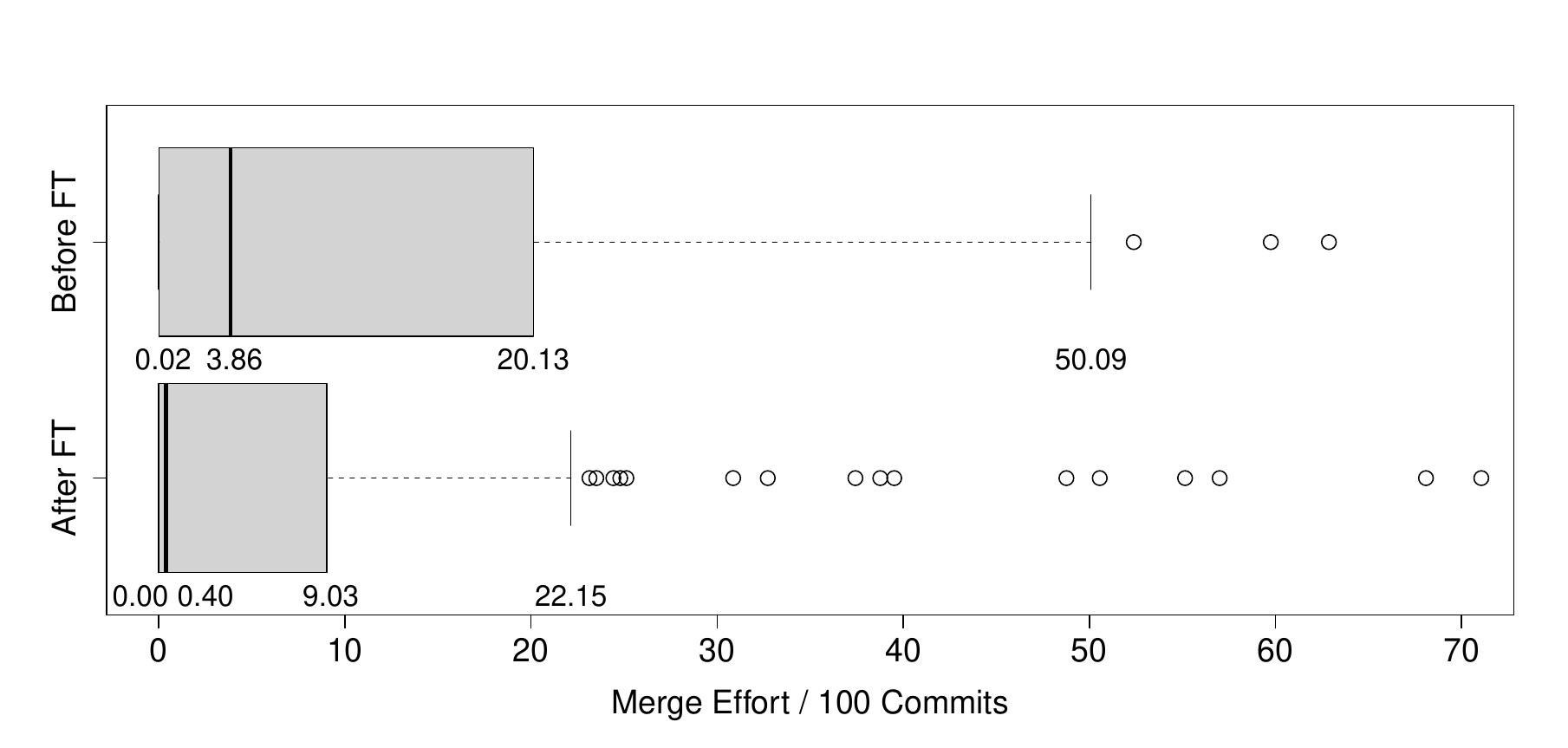}
\end{figure}

Until now, we performed isolated analyses regarding merge, considering two variables: number of merges and effort of each merge. On the one hand, only the effort of each merge presented statistically significant results. On the other hand, both medians have dropped after the adoption of feature toggles. Complementing those analyses, we finish by checking the effects of both variables combined. Thus, using $C_{RQ2}$, we calculated the effort of doing all merges before and after the adoption of feature toggles and normalized the results per 100 commits. In other words, we computed the total merge effort in an interval of 100 commits, which combines both the number of merges in the interval and the effort of each merge. Figure~\ref{fig:fig8} displays a boxplot with the merge effort per 100 commits, before and after the feature toggles adoption. While, on average, the merge effort is around 87.6 lines before adopting feature toggles, this number drops to 18 lines afterward (see Table~\ref{tbl:rq2_summary}).

Proceeding similarly to the previous analyses, we observed non-normality in the data for both groups by applying the Shapiro test ($p\mbox{-}value < 0.001$). Then, we employed the Wilcoxon paired test to check if there is any difference between the data in the two groups. We observed a $p\mbox{-}value = 0.002$, indicating a statistically significant difference. The reduction in the mean and median, as shown in Table~\ref{tbl:rq2_summary}, are around 80\% and 90\%, respectively. We also checked the Cliff’s Delta effect size and observed a small effect size ($0.205$). The interpretation of this result is that there is a significant decrease in the normalized merge effort for the periods before and after the adoption of FT, although the magnitude of this decrease is small.

\begin{figure}
    \centering
     \caption{Comparison of the total merge effort per 100 commits for each programming language. Outliers are omitted for better visualization.}
    \includegraphics[width=1\textwidth]{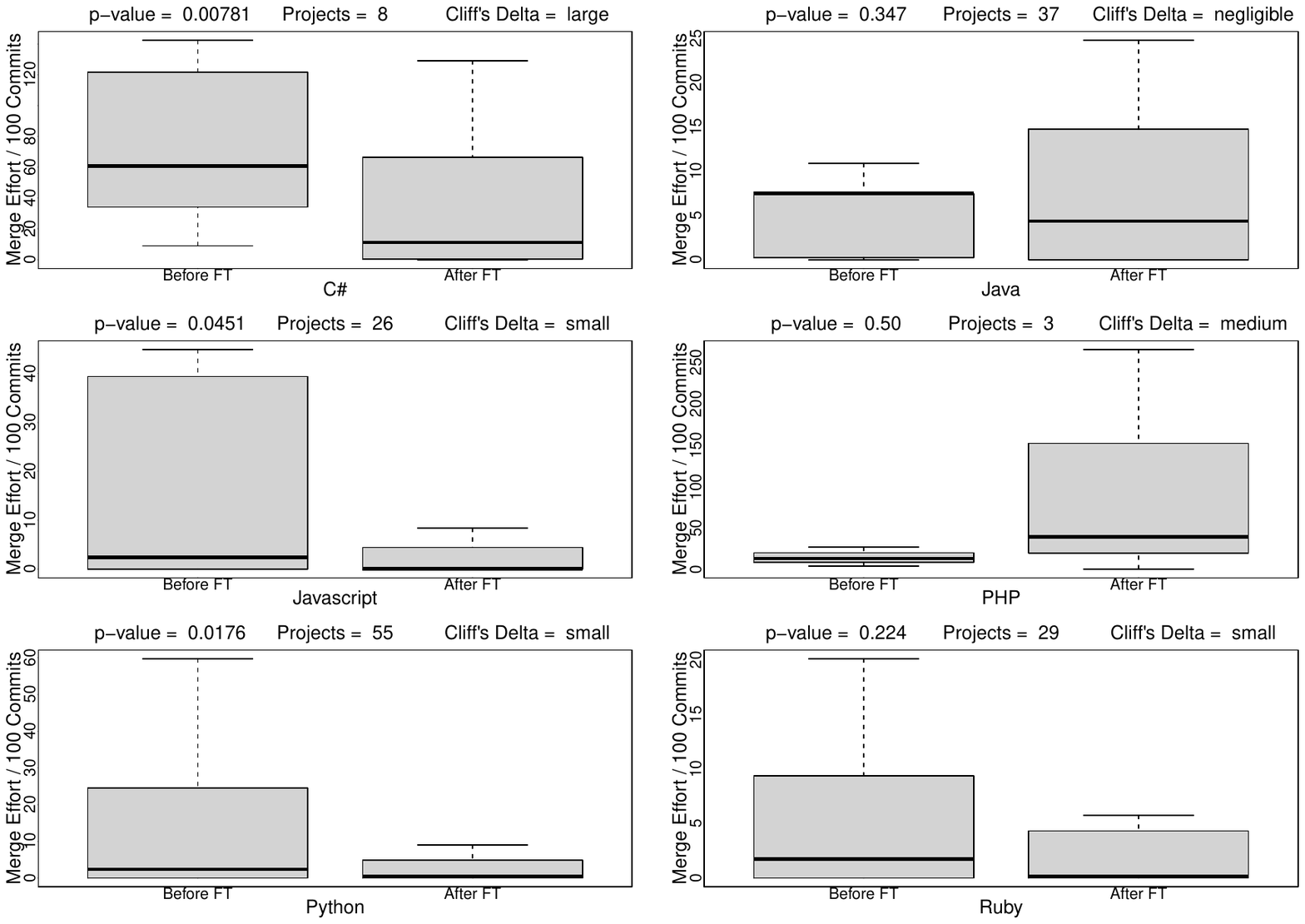}
    \label{fig:boxplot-prog-lang-rq23}
\end{figure}

Figure~\ref{fig:boxplot-prog-lang-rq23} provides an overview of how the total merge effort per 100 commits changed after the adoption of feature toggles for each programming language in $C_{RQ2}$. We checked the normality of the data in each programming language using Shapiro-Wilk's test, but none of them presented a normal distribution, except for PHP. Thus, we applied the nonparametric Wilcoxon paired test \cite{wilcoxon_individual_1945_25} in all cases, as explained in the previous section.  

The total merge effort per 100 commits decreased for all languages, except for PHP. Considering the Bonferroni corrected $\alpha$-value of 0.008, the $H_0$ hypothesis is only rejected for the C\# sample. The p-values suggest that the remaining differences in the samples are not statistically significant. Despite this, the decrease in the measure across the programming languages is compatible with the overall analysis presented in Table~\ref{tbl:rq2_summary}. 

In addition to the median decrease, it is also possible to note in Figure~\ref{fig:boxplot-prog-lang-rq23} that there are smaller variations in the total merge effort per 100 commits values for JavaScript, Python, and Ruby projects after adopting feature toggles (i.e., the distance between the lower and upper limits of the boxplot for the group after FT is much smaller than the group before FT). The opposite happens for Java. For C\#, the variation is almost the same. This variation pattern could also be observed in RQ2.2 (see Figure \ref{fig:boxplot-prog-lang-rq22} in Section~\ref{sec:effort_merge}). We have two possible explanations for this. One is regarding the different frameworks used in each programming language. Some of them may have mechanisms that facilitate the separation of source code, resulting in fewer conflicts and thus, less variation in the merge effort. The other possible explanation is that when using feature toggles, code integration occurs more often (not necessarily branch merges, as we already seen in RQ 2.1) since previously unrelated changes are now performed on the same code. This may result in simple frequent merges. It remains as future work to investigate these hypotheses.

We also analyzed the total merge effort for projects in $C_{Control}$. The Shapiro-Wilk's normality tests suggest that the data before and after the segregation point are not normal, with a $p\mbox{-}value < 2.200 \times 10^{-16}$. We employed the Wilcoxon paired test to check if there is a significant difference between the data in the two groups and found a $p\mbox{-}value = 0.480$, indicating that the difference is not significant.

To summarize, for this research question we could observe a significant difference in the normalized merge effort in $C_{RQ2}$, but not in $C_{Control}$. Thus, this result indicates that the adoption of feature toggles frameworks might have influenced the normalized merge effort decrease in $C_{RQ2}$.

\begin{center}
\noindent\fbox{%
    \parbox{0.985\textwidth}{%
        \textbf{Finding 4}: The normalized effort dedicated to resolving merges showed a statistically significant decrease after the adoption of feature toggles, with a small effect size. This difference was not significant in the control projects. When analyzing the results by programming languages, we observed that the normalized effort after adopting FT has a smaller variance for JavaScript, Python, and Ruby projects. The opposite happened for Java.
    }%
}
\end{center}

\subsection{Do the number of software defects and the time to fix them change after adopting feature toggles? (RQ3)}

In this research question, we aim at checking whether the adoption of feature toggles implies significant changes to the number of defects. We also analyzed the effects of feature toggles in the time needed to fix defects. The analysis procedures for answering all the RQ3 follow, in a similar way, the steps that were taken to answer RQ2. Table \ref{tbl:rq3_summary} displays a summary of the obtained measures for the sub questions in RQ3. The measures obtained for $C_{Control}$ are also displayed for comparison. The results for each sub question are discussed in the next sections.

\begin{table}[]
\begin{center}
\caption{Summary of statistics for sub questions of RQ3 for projects in $C_{RQ3}$ before and after adopting FT, and before and after the segregation point (SP) for projects in $C_{Control}$.}
\label{tbl:rq3_summary}
\begin{tabular}{@{}lllrrrr@{}}
\toprule
\multirow{2}{*}{RQ}  & \multicolumn{1}{c}{\multirow{2}{*}{Measure}}                   & \multirow{2}{*}{}           & \multicolumn{2}{c}{$C_{RQ3}$}               & \multicolumn{2}{c}{$C_{Control}$} \\ \cmidrule(l){4-7} 
                     &                                            &                             & Before FT & After FT                   & Before SP       & After SP        \\ \midrule
\multirow{2}{*}{3.1} & \multirow{2}{*}{Number of defects}         & \multicolumn{1}{l|}{Mean}   & 0.48      & \multicolumn{1}{r|}{2.67}  & 6.50         & 9.67         \\
                     &                                            & \multicolumn{1}{l|}{Median} & 0.03      & \multicolumn{1}{r|}{0.16}  & 1.05         & 1.64         \\
\multirow{2}{*}{3.2} & \multirow{2}{*}{Time per defect}           & \multicolumn{1}{l|}{Mean}   & 14.66      & \multicolumn{1}{r|}{20.79} & 32.66        & 178.51       \\
                     &                                            & \multicolumn{1}{l|}{Median} & 1.56      & \multicolumn{1}{r|}{4.89}  & 17.44        & 75.61        \\
\multirow{2}{*}{3.3} & \multirow{2}{*}{Total time fixing defects} & \multicolumn{1}{l|}{Mean}   & 6.96      & \multicolumn{1}{r|}{25.91} & 189.57       & 3,415.45      \\
                     &                                            & \multicolumn{1}{l|}{Median} & 0.16      & \multicolumn{1}{r|}{0.77}  & 12.15        & 68.78        \\ \bottomrule
\end{tabular}
\end{center}
\end{table}

\subsubsection{Number of defects (RQ3.1)}

\begin{figure}
  \centering
  \caption{Distribution of the normalized number of defects in $C_{RQ3}$, before and after adopting feature toggles. Some outliers were omitted for better visualization. The boxplot shows that the median of the normalized number of defects was 0.03 before the adoption of FT. After the adoption of FT, it increased to 0.16.}
  \label{fig:fig9}
\includegraphics[width=0.7\textwidth]{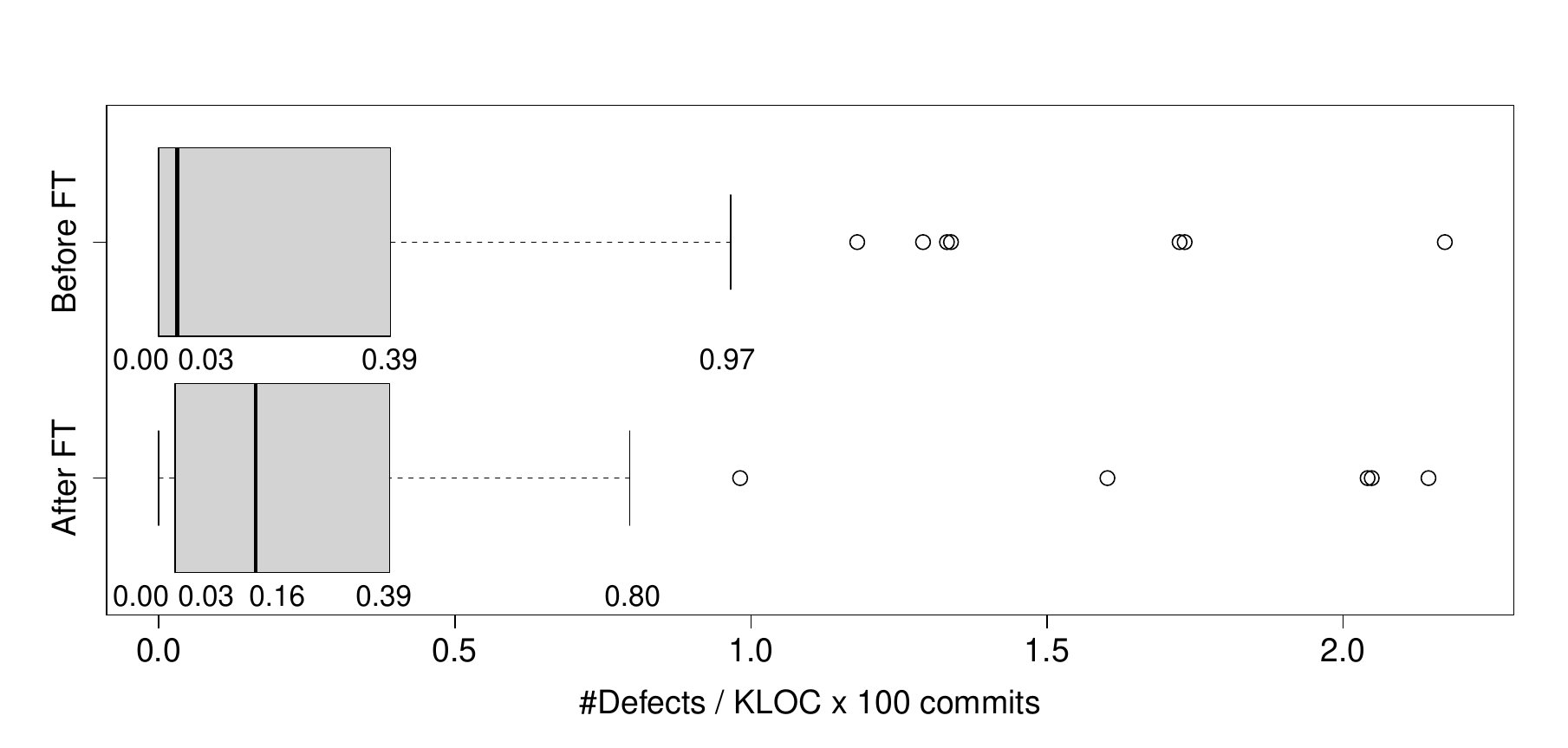}
\end{figure}

We analyze $C_{RQ3}$, which represents the corpus of projects with enough commits for having at least one defect before and after adopting feature toggles. For this research question, we investigate the normalized number of defects in the projects, dividing the available data in two groups, before and after the adoption of feature toggles frameworks. Figure~\ref{fig:fig9} displays the boxplot over the normalized number of defects, before and after feature toggles. 

We first run the Shapiro-Wilk's test to check normality of the number of defects data for both groups. We observed a $p\mbox{-}value < 0.01$ for both distributions, before and after adopting feature toggles. Consequently, we applied the Wilcoxon paired test \cite{wilcoxon_individual_1945_25} and observed $p\mbox{-}value = 0.1184$, not indicating a statistically significant difference.

Table~\ref{tbl:rq3_summary} shows the mean and median number of defects per KLOC in 100 commits. We could observe that both metrics have a substantial increase, by around 456\% and 430\%, respectively, after adopting feature toggles.

Aiming at further understanding why the steep increase in the normalized number of defects was not significant, we also analyzed the number of projects where the normalized number of defects increased and where it decreased. We found that the value increased for 54 projects, and decreased for 25. However, the median increase was 0.05, with a standard deviation of 24.28. On the other hand, the median decrease was 0.22, with a standard deviation of 2.26.

We also performed an analysis of the data segregated by programming language to check whether the observed tendency is uniform. We found that for both Java and C\#, the likelihood of increasing the number of defects after the adoption of feature toggles increases by 35.8\% and 46.3\%, respectively. Moreover, all five C\# projects had the number of defects increased after adopting feature toggles. For the remaining programming languages, the tendency is to decrease the number of defects after adopting feature toggles. We could also observe that PHP has all its two projects with the number of defects decreasing.

When analyzing the boxplots in Figure \ref{fig:fig9}, we can observe that although the median from "\textit{before FT}" and "\textit{after FT}" are far from each other, the Q3 of the boxplot is practically the same. By looking at the values above Q3, we can observe that "\textit{before FT}" has higher values, higher outliers, and a higher upper limit. On the other hand, looking at the values below Q3, "\textit{before FT}" has about 25\% of the projects with a zero value for the metric, since Q1 is not visible. Furthermore, 50\% of the "\textit{before FT}" values are very low ($< 0.03$), coinciding with Q2 (the 25\% lowest values) for the "\textit{after FT}" sample. Hence, we conclude that "\textit{before FT}" has a long tail distribution with a big variance, with many very low and some very big values. On the other hand, "\textit{after FT}" has less variance and not many projects with a zero value. This probably helped to shift the median. However, the fact that the behavior is the same in Q3 and inverse in Q4 explains why it was not possible to reject $H_{0}$.

We also analyzed if there was a significant difference for the normalized number of defects in $C_{Control}$. First we checked normality using Shapiro-Wilk's test. We found $p\mbox{-}value = 9.031 \times 10^{-16}$ for the number of defects in the first half of the history and $p\mbox{-}value < 2.200 \times 10^{-16}$ for the number of defects in the last half of the history. This suggests that the number of defects data do not follow a normal distribution. Thus, we employed the nonparametric Wilcoxon paired test and found a $p\mbox{-}value = 0.060$, which suggests that there is no significant difference between the two groups of data. However, the increase of about 48\% in the mean, and 56\% in the median normalized number of defects in $C_{Control}$ is much smaller than the increase observed in $C_{RQ3}$. Considering that we also did not find a significant difference for $C_{RQ3}$, we cannot confirm that the adoption of feature toggles frameworks had any influence on the normalized number of defects.

\begin{center}
\noindent\fbox{%
    \parbox{0.985\textwidth}{%
        \textbf{Finding 5}: We could not observe a statistically significant difference in the normalized number of defects after adopting feature toggles. However, the mean and median normalized number of defects have increased by more than 400\%. Meanwhile, there was also no significant difference in the control projects, but they showed a much smaller increase, by about 48\% and 56\% in the mean and the median, respectively.
    }%
}
\end{center}

\subsubsection{Time per defect (RQ3.2)}

\begin{figure}
  \centering
  \caption{Distribution of time (in days) needed to fix a defect in $C_{RQ3}$, before and after the adoption of feature toggles. Some outliers were omitted to ease visualization. The boxplot shows that before adopting FT, projects needed a median of 1.6 days to fix a defect. After the adoption of FT, this median time increased to 4.9 days.}
  \label{fig:fig10}
\includegraphics[width=0.7\textwidth]{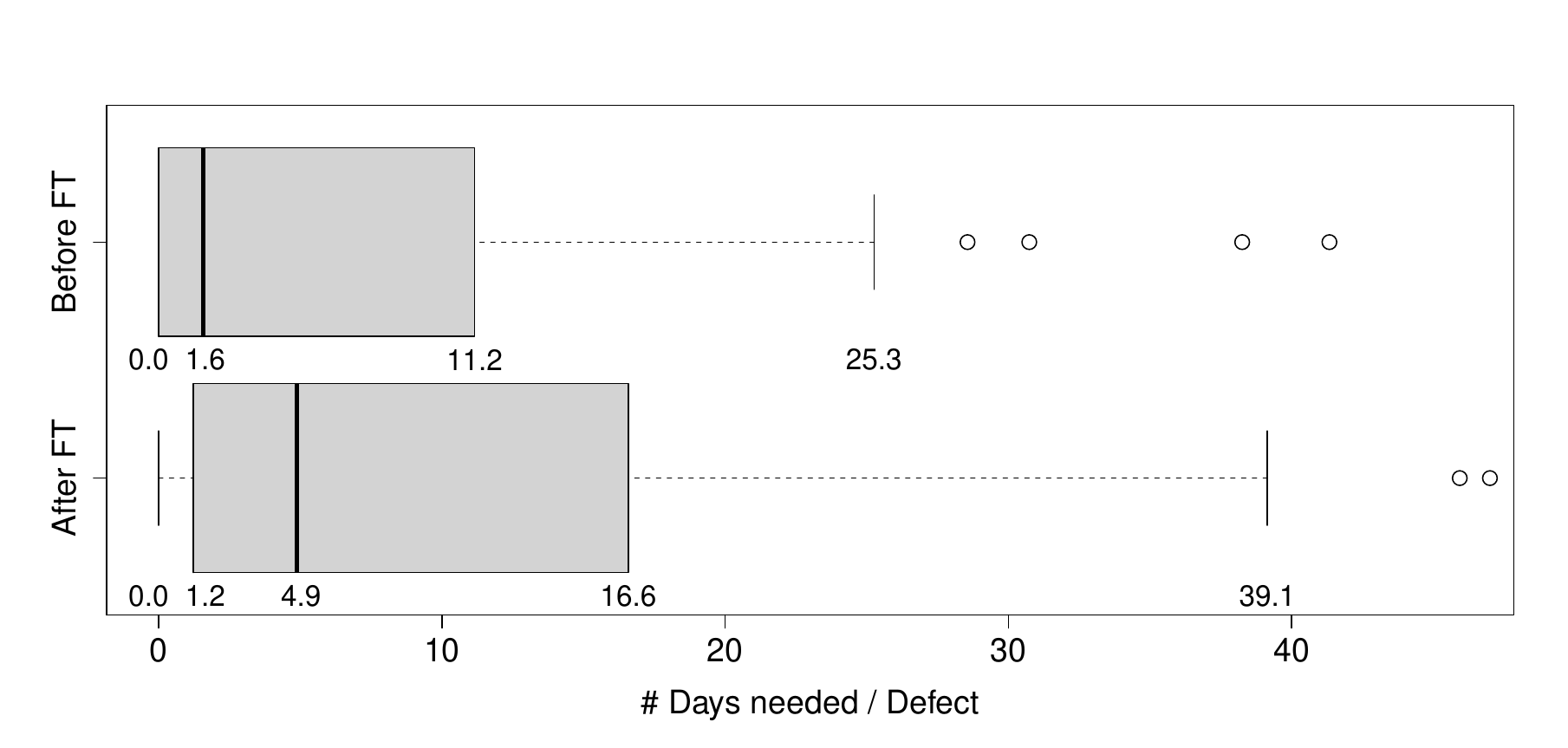}
\end{figure}

We verified whether the amount of time (measured in days) needed to fix a defect has significantly changed after adopting feature toggles. We run this analysis over $C_{RQ3}$, which contains all projects with at least one defect, and enough commits to have at least one defect before and one defect after the adoption of feature toggles, except for outliers. Figure~\ref{fig:fig10} shows the boxplot of the number of days needed to fix a defect, before and after the introduction of feature toggles. 

Again, we checked normality using the Shapiro-Wilk's test and found $p\mbox{-}value = 1.39 \times 10^{-15}$ for the data before the adoption of FT and $p\mbox{-}value < 2.2 \times 10^{-16}$ for the data after the adoption of FT. Since the distribution of the data for both groups is not normal, we applied the Wilcoxon paired test to check if there is a significant difference between them. We observed a $p\mbox{-}value = 0.097$, indicating that the difference between the amount of time for fixing defects before and after adopting FT frameworks is not statistically significant. We also checked the Cliff’s Delta and observed a small effect size ($-0.292$).

Closing this sub-question, in Table~\ref{tbl:rq3_summary}, we show the average amount of time needed to fix a defect. We could observe an increase of about 42\% in the mean, and 213\% in the median.

We also analyzed this variable on $C_{Control}$. We first checked the normality of the data using Shapiro-Wilk's test. We found a $p\mbox{-}value = 3.411 \times 10^{-15}$ for the data in the first half of the history and $p\mbox{-}value = 5.879 \times 10^{-16}$ for the data in the last half of the history. This result suggests that the distributions are not normal. Thus, we used the nonparametric Wilcoxon paired test to check if there are differences between the data in both groups. We found a $p\mbox{-}value = 7.663 \times 10^{-10}$, which suggests that there is a significant difference. We further investigated this difference by calculating Cliff's Delta and found $d = -0.570$, which suggests that the effect size is large. In fact, according to Table~\ref{tbl:rq3_summary}, the mean time to fix a defect in $C_{Control}$ has increased about 446\% and the median about 333\%.

We found a significant difference in the time to fix a defect in $C_{Control}$, but not in $C_{RQ3}$. This finding suggests that, overall, the time to fix a defect increases, but since this increase was not significant in $C_{RQ3}$, the adoption of feature toggles frameworks might have influenced this variable.

\begin{center}
\noindent\fbox{%
    \parbox{0.985\textwidth}{%
        \textbf{Finding 6}: We could not observe a statistically significant difference between the time needed to fix a defect before and after the adoption of feature toggles frameworks. Both the mean and the median time to fix a defect have increased. On the other hand, we found a significant increase for this measure in the control projects. This suggests that the time to fix a defect naturally grows and the adoption of feature toggles frameworks might have changed this behavior in the analyzed projects. 
    }%
}
\end{center}

\subsubsection{Total time fixing defects (RQ3.3)}

\begin{figure}
  \centering
  \caption{Distribution of normalized time (in days) to fix defects in $C_{RQ3}$, before and after adopting feature toggles. Some outliers were omitted for better visualization. The boxplot shows that the median of the normalized time to fix a defect was 0.16 days before the adoption of FT. After the adoption of FT, it increased to 0.77 days.}
  \label{fig:fig11}
\includegraphics[width=0.7\textwidth]{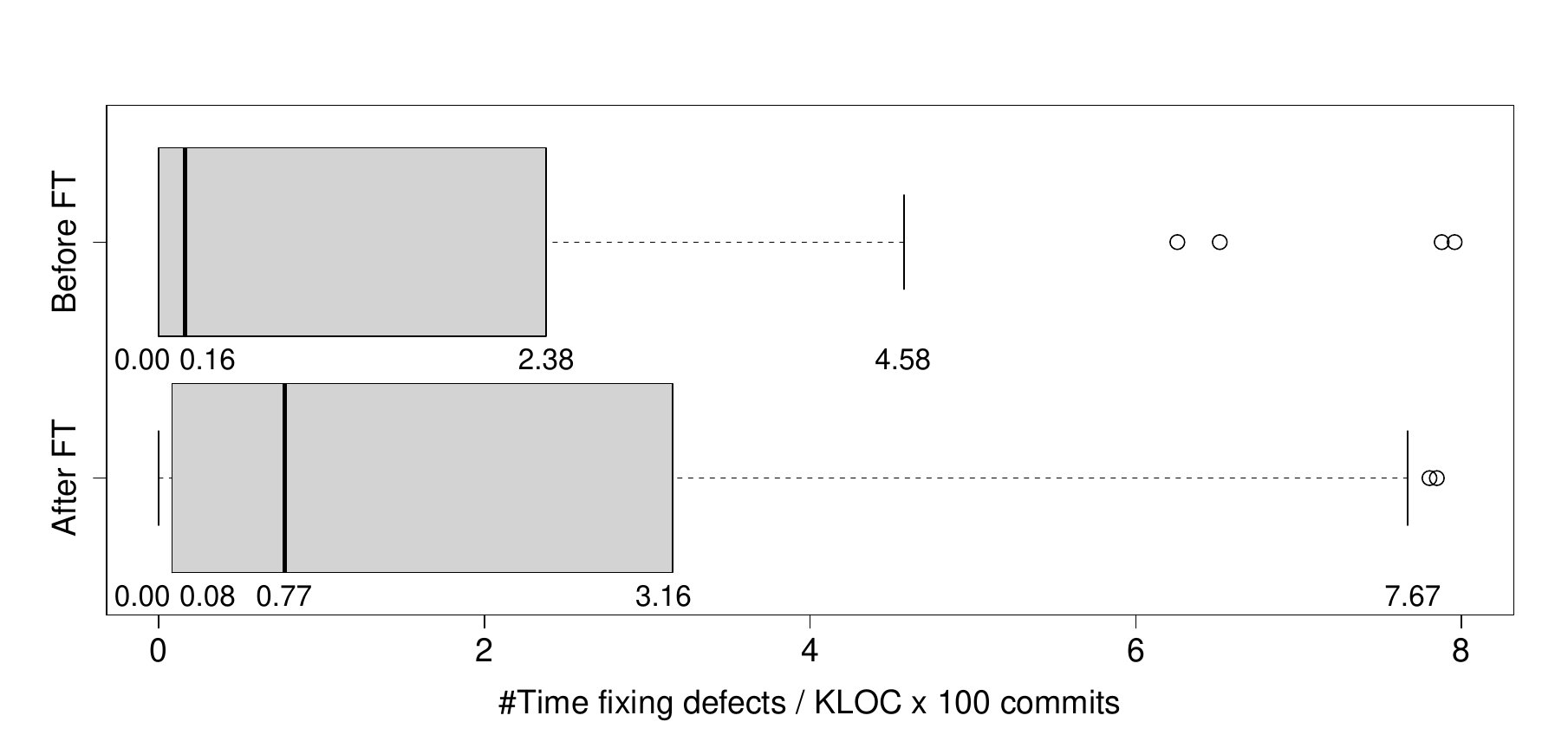}
\end{figure}

Complementing the previous analyses, we verified whether the normalized amount of time (measured in days) needed to fix defects has statistically significant changes after adopting feature toggles. Similar to the analysis presented in Section~\ref{sec:total_merge_effort}, in this analysis, we assessed the combination of the number of defects and the time needed to fix them together. Figure~\ref{fig:fig11} shows the boxplot of the normalized time (in days) needed to fix defects before and after the introduction of feature toggles. In other words, it shows how many days were needed to fix all defects that occurred per KLOC in an interval of 100 commits, before and after the adoption of feature toggles. 

Again, we verified the normality of the data using the Shapiro-Wilk's test ($p\mbox{-}value < 0.001$) and applied Wilcoxon paired test. We observed a $p\mbox{-}value = 0.1661$, not indicating a statistically significant difference in the amount of time for fixing defects. We also checked the Cliff’s Delta effect size and observed a medium effect size ($0.2361$). 

Table~\ref{tbl:rq3_summary} shows the mean and median normalized time to fix defects. After adopting feature toggles, the average time to fix defects per KLOC in 100 commits increased by around 272\%, and the median time increased by almost 381\%.

Aiming at further understanding why the increase in the normalized time to fix defects was not significant, we also analyzed the number of projects where the values increased and where it decreased. We found that it increased for 53 projects, decreased for 23, and stayed the same for three projects. The median increase was 0.49, with a standard deviation of 216.05. On the other hand, the median decrease was 5.95, with a standard deviation of 28.45.

We also analyzed the data segregated by programming language. We found that the likelihood of increasing the time to fix defects for Java and C\# increases by 27.8\% and 49\%, respectively. Moreover, all five C\# projects had the time to fix defects increased after adopting feature toggles. For PHP, Python, and Ruby, the tendency was to decrease the time to fix defects after adopting feature toggles, with likelihoods increased by 71.7\%, 22.6\%, and 37.3\%, respectively. An interesting observation is that for JavaScript and Ruby, the likelihood of the time to fix defects staying the same increases by 251.1\% and 75.5\%, respectively.

Similar to how we analyzed the boxplot for the normalized number of defects, we now analyze for the normalized time to fix defects. When analyzing the boxplots in Figure \ref{fig:fig11}, we can observe that the distance between the medians from "\textit{before FT}" and "\textit{after FT}" is similar to the distance between their upper limit (Q3). However, this does not apply to their lower limit, as 50\% of the values in "\textit{before FT}" are very low ($<0.16$). It can also be noted that despite the upper limit value of "\textit{before FT}" being lower than the one in the "\textit{after FT}" sample, it has higher outliers. From this analysis, we can conclude that "\textit{before FT}" has smaller values and smaller variance than "\textit{after FT}". Nonetheless, it also has bigger outliers. Overall, it may not have been possible to reject $H_{0}$ due to this opposite behavior, which also happened in RQ3.1.

Finally, we also performed the analysis of the total time fixing defects in $C_{Control}$. We found that the distribution of the data in the first half of the history is not normal with a $p\mbox{-}value = 1.676 \times 10^{-15}$. The same was found for the distribution for the data in the last half of the history, with a $p\mbox{-}value < 2.200 \times 10^{-16}$. We used Shapiro-Wilk's test for both of them. Since they are not normal, we used the Wilcoxon paired test to check if their difference is significant. We found a $p\mbox{-}value = 1.636 \times 10^{-06}$, which suggests that there is a significant difference. We also calculated Cliff's Delta to check the magnitude of the difference. We found $d = -0.351$, which suggests that the effect size is medium.

Similarly to the time to fix defects analysis, we did not find a significant difference for the total time fixing defects in $C_{RQ3}$. However, this difference was significant in $C_{Control}$, with the total time fixing defects increasing. This suggests that this variable naturally grows in the control projects. Since this increase was not significant in $C_{RQ3}$, this result indicates that the adoption of feature toggles frameworks might have influenced this behavior change.

\begin{center}
\noindent\fbox{%
    \parbox{0.985\textwidth}{%
        \textbf{Finding 7}: We could not observe a statistically significant increase in the normalized time needed to fix defects after adopting feature toggles. However, we could observe a medium effect size and an increase of 381\% in the median. We could also observe a medium effect size in the control projects, with a significant difference. This suggests that the adoption of feature toggles frameworks might have influenced the behavior change in the total time fixing defects. 
    }%
}
\end{center}

\section{Threats to Validity}\label{sec:threats}

Although we aimed at minimizing the threats to the validity of our study, some decisions may have affected the results, as discussed in the following. 

To compose the analysis corpus, we searched for open-source projects based on a heuristic that checks whether the projects use any feature toggles frameworks. Our first step in this direction was the identification of feature toggles frameworks. Although we did our best to find frameworks listed in specialized websites, books, and repositories, we may have missed some specific feature toggle frameworks. The potential consequence of such a threat is the absence of some relevant projects in $C_{Initial}$ (false negatives). Nonetheless, due to the extensive search process adopted in our research to find frameworks, we believe that the most relevant frameworks were included. Next, we used some specific keywords to identify their use, mainly based on “imports”. This heuristic may incur in false positives for projects that import a feature toggles framework but do not use it extensively throughout their histories. We checked a sample of 849 of the 949 projects from $C_{RQ1}$ and observed that only one project stopped using a feature toggles framework later in history after adopting it. \footnote{We did not check all of the 949 projects because since this analysis was made after the data collection, some of the data was not available on GitHub anymore.} Thus, we expect the impact of false positives to be very low. Additionally, the heuristic can also incur in false negatives for projects that use feature toggles but do not employ any framework. However, these projects are not the focus of our paper, as we are concerned with projects that adopt a framework.

The definition of a threshold for the minimum number of commits before and after adopting feature toggles may have included irrelevant projects or excluded relevant projects from $C_{RQ1}$. To mitigate this threat, we adopted Tukey's Fence formula to define the threshold. This technique provided us a more reliable threshold, based on the minimum number of commits that guarantees that all projects, but outliers, have at least one merge or defect.

Rebase is a mechanism available on \textit{Git} that allows a developer to reapply commits on top of others. Thus, a rebased project history looks like it was linear, commit after commit, although it might not have been. Some developers may opt to use rebases instead of merges. In this paper, we focus only on merges, since in general, it may not be possible to identify if and when a project has used rebases. We found a study that investigates rebases on open-source projects, but the technique they use is limited to pull requests only\cite{merges_rebases}. Using this technique in our analysis would limit our dataset even more. Thus, we point out as future work to investigate the relationship between the use of rebases and feature toggles.

We adopted a heuristic that was already used by other studies to estimate the number of branch merges on each project's commit history. As previously described, it classifies a merge commit as a branch merge if more than one unique developer has participated on each side of the merge. This heuristic may be subject to false positives in rare situations where a developer commits with different credentials in the same branch, and the project does not have a .mailmap file unifying such credentials. Additionally, it may be susceptible to false negatives for simple named branches, when only one developer has committed to the branch. To mitigate these problems, we complemented the heuristic with an analysis of the merge commit message, checking if it indicates a branch merge. Nevertheless, the situations that may lead to false positives are quite uncommon, and the effect of a residual false negative would not affect the soundness of the corpus, because it just misses the opportunity of analyzing a valid merge.

When creating $C_{RQ3}$ for defect analysis, we used keywords to search for issues that represent defects. Our heuristic search in the labels, title, and description of the issues for specific keywords. Although we took care to select appropriate keywords, we may have missed some important keywords or added some inappropriate keywords by mistake. However, such error would affect both samples (before and after adopting feature toggles) in an equivalent intensity. Moreover, we performed a sensitivity analysis using 10 random issues from 10 random projects from $C_{RQ1}$, summing up a total of 100 random issues. One author manually classified these issues, and another author revised such classification. When a divergence was found, a third author helped to reach a decision. After this classification, we compared our results with the results obtained by using the keywords approach. Out of the 100 issues, the keywords classified 77 correctly and 23 incorrectly, showing an accuracy of 77\%. Among the 23 incorrect classifications, we have 8 false positives and 15 false negatives, leading to a precision of 68\% and a recall of 53\%. These numbers show that our conclusions should be considered with care. However, this is not equivalent to random guessing, as the number of true positives (17) is small compared to the number of true negatives (60). Random guessing in this sample would lead to a precision of 32\% and a recall of 50\%.

Finally, although we have normalized our data to protect against two confounding factors – the number of commits and the size of the project before and after adopting feature toggles – it may still be exposed to other confounding factors. A potential confounding factor, regarding the time to fix defects, is the number of developers. On the one hand, the more the number of developers, the more the discussions on each defect. On the other hand, the more the number of developers, the faster the coding of patches. To mitigate this threat, we replicated our study using 43 projects from $C_{Cleaned}$. Those projects have at least one merge before and one merge after the adoption of FT, and at least one reported defect before and one reported defect after the adoption of FT. We normalized the dependent variables by the number of developers, besides the number of commits and the size. We observed the same tendency of increasing the normalized time needed to fix defects, but with negligible effect size and without statistical significance.

\section{Related Work}\label{sec:related}

Although feature toggle is gaining attention from major software companies, few existing studies provide evidence about the benefits and limitations of using it. In this section, we present studies that shed light on the benefits and limitations of feature toggles based on empirical findings, theoretical analyses, case studies, or literature and practitioner surveys. 

Neely and Stolt~\cite{neely_continuous_2013_8} describe the results of changes implementing the continuous delivery process in a specific software company. Due to the effort of merging long-running branches and integration delays imposed by feature branches, the studied company started to practice trunk-based development with the adoption of a feature toggles framework. Similarly, Rehn~\cite{rehn_continuous_2012_9} highlights the importance of continuous integration to reduce technical problems and detect defects early. The author compared some collaborative development techniques and suggested using feature toggles for continuous integration instead of feature branches. Both studies do not provide quantitative evidence for their recommendation. The results of our work somehow contrast with their suggestions, considering that we could not observe significant changes in the number of branch merges after the adoption of feature toggles (Finding 2). Furthermore, although they highlight the importance of feature toggles for continuous integration to detect defects early, we observed that the number of defects increased after adopting feature toggles (Finding 5).

Rahman et al.~\cite{rahman_feature_2016_4} report the results of using feature toggles in the Google Chrome project. They compared the development of Google Chrome before and after adopting feature toggles. Before Google adopted feature toggles in Chrome, the development team usually worked in iterations of 6 weeks on a single release branch. Moreover, developers committed their changes directly to this branch. Hence, this release branch was blocked until all features were finished, causing a considerable effort for the developers to stabilize the code (merging 500 patches), introducing delays to meet deadlines and to fix defects. After adopting feature toggles, they could reduce the total merge effort by making the merge more predictable. Moreover, they reported that developers could fix defects with less effort by avoiding the need to switch branches, losing uncommitted changes. Although their study considered only one project, we could observe a similar result in a much bigger corpus: a substantial decrease in the overall merge effort (Finding 4). However, when considering the increase in the number of defects (Finding 5), we could observe a severe increase in the overall time to fix them (Finding 7). 

Additionally, Rahman et al.~\cite{rahman_feature_2016_4} analyzed the life cycle of feature toggles in the Google Chrome project and highlighted the need to control the toggles debt through a disciplined and proactive feature design. Similarly, Bird~\cite{bird_feature_2014_28} discusses the advantages and disadvantages of using feature toggles, emphasizing the need for short-lived feature toggles and the need for toggle debt control due to undesired behavior of features and unpredictable results (defects). In our study, we could observe an increase in defects (Finding 7), which might be a consequence of inappropriate control of toggle debt. The results presented by Mahdavi-Hezaveh et al.~\cite{mahdavi2021software} seem to reinforce this. They performed a survey with 20 practitioners from at least 17 different companies that use feature toggles. They found that, although all of them use a dedicated tool to create and manage feature toggles, the least used practice is the clean-up of unused feature toggles. In this sense, Ramanathan et al.~\cite{ramanathan2020piranha} developed a tool called Piranha to semi-automatically deal with stale feature toggles at Uber. The tool searches the source code for stale feature flags, performs an automatic refactoring to clean-up the source code, and then assigns a developer to review the changes. According to them, the tool was used for about one year and five months and helped 200 developers to delete 1,381 stale flags. From these stale flags, 65\% did not need any manual change to be applied.

In more recent work, Rahman et al.~\cite{rahman_modular_2018_6} analyze the software architecture of Google Chrome by extracting feature toggles from the code and identifying the relationship between the toggles. Thus, they could map all used features into a modular representation to create a feature toggles architectural view of Google Chrome. Throughout this study, they showed how the feature toggle view could give new perspectives into the feature architecture of a system.

Schermann et al.~\cite{schermann_empirical_2016_7} present a survey with developers of software companies, characterizing the profiles and the main development techniques adopted by those companies for continuous delivery and deployment. Although companies like Google and Facebook have adopted feature toggles instead of branches for collaborative software development, this survey shows that most companies still do not consider using feature toggles due to extra complexity in the code. Our study complements theirs by providing evidence about the pros and cons of replacing branches with feature toggles. This evidence may help companies on a conscientiously move to feature toggles.

Meinicke et al.~\cite{meinicke2020exploring} interviewed nine feature toggle experts and correlated their answers with existing literature to identify commonalities and differences between configuration flags and feature toggles. They found that they are similar concepts but have different characteristics and requirements. Despite this, they highlight that there is space to transfer existing knowledge from the configuration flags field, which exists for a long time, to feature toggles.

In another paper, Meinicke et al.~\cite{meinicke2020capture} proposed a heuristic for identifying the use of feature toggles, by analyzing project commit messages and comparing them to a set of regular expressions. They employ this heuristic on a preliminary empirical study to select projects from GitHub that use feature toggles. Based on these projects, they manually select 100 of them to analyze how frequently feature toggles are removed after being included in the projects. They found that the majority of the projects often clean up existing toggles, but some of them remain in the source-code for a long time. Our study was performed in parallel to theirs. Different from them, we identified the use of feature toggles on open-source projects by analyzing the import of related frameworks/libraries. In the future, our study can be replicated using their approach to check how the results compare.

Overall, none of the previous works provides quantitative evidence, based on a large corpus, about the effects of feature toggles in the number of branch merges and the necessary effort to resolve them. Additionally, we could not find any related work that studies the effects of feature toggles in the number of reported defects and the necessary time to fix them.

\section{Conclusions}\label{sec:conclusions}

In this work, we studied the effects of applying feature toggles on 949 open-source projects written in 6 different programming languages. We first identified the moment in which each project adopted feature toggles. Then, we observed if the number and effort of branch merges changed after the adoption, as well as the number of defects and the average time to fix them. The corpora and scripts used in the analyses are available in our companion site: \url{https://gems-uff.github.io/feature-toggles}.

Surprisingly, the adoption of feature toggles did not lead to statistically significant changes in the number of branch merges. We could observe in our study that some projects, in fact, completely migrated to trunk-based development after the adoption of feature toggles. However, in this paper, we focused on analyzing the results more quantitatively. It remains as future work a qualitative investigation of the reasons why open-source projects adopt feature toggles. If, on the one hand, the number of merges did not change, on the other hand, the total merge effort dropped significantly on average (80\%). This result indicates that feature toggles may become an alternative to branches in collaborative software development, potentially reducing the risk of broken features or undesirable behavior due to the merge process. Besides, it may also reduce developers' pain of performing complicated merges. However, this hypothesis needs to be further investigated by future studies.

Although we could not observe a statistically significant difference in either the number of defects and time to fix them, the mean number of defects increased by 456\%, the mean time to fix a defect increased by 42\%, and the mean total time fixing defects increased by 272\%. This result is aligned with the indication of some authors that the feature toggles technique may lead to a growth of application defects. Despite this, we could also observe an increase in these measures in the control projects, which suggests that this might happen due to factors other than the use of feature toggles frameworks. We suggest replication of this study over a more extensive corpus in the future.

As future work, we intend to investigate the stabilization of the merge effort for some programming languages that we found in RQ2.3 (Section \ref{sec:total_merge_effort}). We also intend to study how the number of developers in projects that adopted feature toggles could affect the branching merge process. We also want to study whether the increase in the number of defects is related to the additional test complexity imposed by feature toggles. This study could help devise new approaches to plan and conduct tests in the context of feature toggles. 

Finally, we want to study a corpus composed of projects that use feature toggles since their creation, checking whether we could observe the same results regarding merge and defects.

\section*{Acknowledgements}

We would like to thank CAPES, CNPq, and FAPERJ for the financial support.

\bibliographystyle{plainnat}   
\bibliography{references}  

\begin{thebibliography}{39}
\providecommand{\natexlab}[1]{#1}
\providecommand{\url}[1]{\texttt{#1}}
\expandafter\ifx\csname urlstyle\endcsname\relax
  \providecommand{\doi}[1]{doi: #1}\else
  \providecommand{\doi}{doi: \begingroup \urlstyle{rm}\Url}\fi

\bibitem[Adams and McIntosh(2016)]{adams2016modern}
Bram Adams and Shane McIntosh.
\newblock Modern release engineering in a nutshell--why researchers should
  care.
\newblock In \emph{2016 IEEE 23rd international conference on software
  analysis, evolution, and reengineering (SANER)}, volume~5, pages 78--90.
  IEEE, 2016.

\bibitem[Barnett and Lewis(1994)]{barnett_outliers_1994_21}
Vic Barnett and Toby Lewis.
\newblock \emph{Outliers in {Statistical} {Data}}.
\newblock Wiley, USA, 3rd edition, April 1994.
\newblock ISBN 0-471-93094-6.

\bibitem[Berczuk(2003)]{berczuk_pragmatic_2003_10}
S.~Berczuk.
\newblock Pragmatic {Software} {Configuration} {Management}.
\newblock \emph{IEEE Software}, 20\penalty0 (2):\penalty0 15--17, 2003.
\newblock ISSN 0740-7459.
\newblock \doi{http://dx.doi.org/10.1109/MS.2003.1184160}.

\bibitem[Bird(2014)]{bird_feature_2014_28}
Jim Bird.
\newblock Feature {Toggles} are one of the {Worst} kinds of {Technical} {Debt}
  - {DZone} {DevOps}.
\newblock \url{https://dzone.com/articles/feature-toggles-are-one-worst},
  November 2014.
\newblock URL \url{https://dzone.com/articles/feature-toggles-are-one-worst}.

\bibitem[Bonferroni(1936)]{bonferroni_teoria_1936_26}
C.~Bonferroni.
\newblock Teoria statistica delle classi e calcolo delle probabilita.
\newblock \emph{Pubblicazioni del R Istituto Superiore di Scienze Economiche e
  Commericiali di Firenze}, 8:\penalty0 3--62, 1936.

\bibitem[Costa et~al.(2014)Costa, Figueiredo, Ghiotto, and
  Murta]{costa_characterizing_2014_17}
C.~Costa, J.~J.~C. Figueiredo, G.~Ghiotto, and L.~Murta.
\newblock Characterizing the {Problem} of {Developers}' {Assignment} for
  {Merging} {Branches}.
\newblock \emph{International Journal of Software Engineering and Knowledge
  Engineering}, 24\penalty0 (10):\penalty0 1489--1508, December 2014.
\newblock ISSN 0218-1940.
\newblock \doi{10.1142/S0218194014400166}.
\newblock URL
  \url{http://www.worldscientific.com/doi/abs/10.1142/S0218194014400166}.

\bibitem[Costa et~al.(2016)Costa, Figueiredo, Murta, and
  Sarma]{costa_tipmerge_2016_18}
Catarina Costa, Jair Figueiredo, Leonardo Murta, and Anita Sarma.
\newblock {TIPMerge}: {Recommending} {Experts} for {Integrating} {Changes}
  {Across} {Branches}.
\newblock In \emph{Proceedings of the 2016 24th {ACM} {SIGSOFT} {International}
  {Symposium} on {Foundations} of {Software} {Engineering}}, {FSE} 2016, pages
  523--534, New York, NY, USA, 2016. ACM.
\newblock ISBN 978-1-4503-4218-6.
\newblock \doi{10.1145/2950290.2950339}.
\newblock URL \url{http://doi.acm.org/10.1145/2950290.2950339}.

\bibitem[Dias et~al.(2020)Dias, Borba, and Barreto]{dias2020understanding}
Klissiomara Dias, Paulo Borba, and Marcos Barreto.
\newblock Understanding predictive factors for merge conflicts.
\newblock \emph{Information and Software Technology}, 121:\penalty0 106256,
  2020.

\bibitem[DigitalOcean(2018)]{digitalocean_developers_2018_12}
DigitalOcean.
\newblock Developers are using {CI} more than {CD}, report finds.
\newblock \url{https://sdtimes.com/cicd/developers-using-ci-cd-report-finds/},
  March 2018.

\bibitem[Feitelson et~al.(2013)Feitelson, Frachtenberg, and
  Beck]{feitelson_development_2013_2}
Dror~G. Feitelson, Eitan Frachtenberg, and Kent~L. Beck.
\newblock Development and deployment at facebook.
\newblock \emph{IEEE Internet Computing}, 17\penalty0 (4):\penalty0 8--17,
  2013.

\bibitem[Fowler(2010)]{fowler_bliki_2010_13}
Martin Fowler.
\newblock bliki: {FeatureToggle}.
\newblock \url{https://martinfowler.com/bliki/FeatureToggle.html}, October
  2010.

\bibitem[Heys(2014)]{heys_alm_nodate_3}
Bill Heys.
\newblock {ALM} {Rangers} - {Software} {Development} with {Feature} {Toggles}.
\newblock
  \url{https://docs.microsoft.com/en-us/archive/msdn-magazine/2014/may/alm-rangers-software-development-with-feature-toggles},
  2014.
\newblock Library Catalog: docs.microsoft.com.

\bibitem[Hodgson(2016)]{pete_hodgson_feature_2016_14}
Pete Hodgson.
\newblock Feature {Toggles}.
\newblock \url{https://martinfowler.com/articles/feature-toggles.html}, August
  2016.

\bibitem[Httermann(2012)]{httermann_devops_2012_16}
Michael Httermann.
\newblock \emph{{DevOps} for {Developers}}.
\newblock Apress, New York, October 2012.
\newblock ISBN 978-1-4302-4570-4.

\bibitem[Hubaux et~al.(2012)Hubaux, Jannach, Drescher, Murta, Männistö,
  Czarnecki, Heymans, Nguyen, and Zanker]{hubaux_unifying_2012_15}
Arnaud Hubaux, Dietmar Jannach, Conrad Drescher, Leonardo Murta, Tomi
  Männistö, Krzysztof Czarnecki, Patrick Heymans, Tien Nguyen, and Markus
  Zanker.
\newblock Unifying {Software} and {Product} {Configuration}: {A} {Research}
  {Roadmap}.
\newblock In \emph{Proceedings of the 2012 {International} {Conference} on
  {Configuration} - {Volume} 958}, {CONFWS}'12, pages 31--35, Aachen, Germany,
  Germany, 2012. CEUR-WS.org.
\newblock URL \url{http://dl.acm.org/citation.cfm?id=3053577.3053583}.

\bibitem[Ji et~al.(2020)Ji, Chen, Yi, and Mao]{merges_rebases}
Tao Ji, Liqian Chen, Xin Yi, and Xiaoguang Mao.
\newblock Understanding merge conflicts and resolutions in git rebases.
\newblock In \emph{2020 IEEE 31st International Symposium on Software
  Reliability Engineering (ISSRE)}, pages 70--80, 2020.
\newblock \doi{10.1109/ISSRE5003.2020.00016}.

\bibitem[Knuth(1997)]{artofprogramming}
Donald~E. Knuth.
\newblock \emph{The Art of Computer Programming, Volume 2 (3rd Ed.):
  Seminumerical Algorithms}.
\newblock Addison-Wesley Longman Publishing Co., Inc., USA, 1997.
\newblock ISBN 0201896842.

\bibitem[Kumar(2017)]{kumar_dilemma_2017_5}
Kshitij Kumar.
\newblock \emph{Dilemma of speed vs. scale in software system development best
  practices from industry leaders}.
\newblock Thesis, Massachusetts Institute of Technology, 2017.
\newblock URL \url{http://dspace.mit.edu/handle/1721.1/110137}.

\bibitem[Macbeth et~al.(2011)Macbeth, Razumiejczyk, and
  Ledesma]{macbeth_cliffs_2011_23}
Guillermo Macbeth, Eugenia Razumiejczyk, and Rubén~Daniel Ledesma.
\newblock Cliff's {Delta} {Calculator}: {A} non-parametric effect size program
  for two groups of observations.
\newblock \emph{Universitas Psychologica}, 10\penalty0 (2):\penalty0 545--555,
  May 2011.
\newblock ISSN 1657-9267.
\newblock URL
  \url{http://www.scielo.org.co/scielo.php?script=sci_abstract&pid=S1657-92672011000200018&lng=en&nrm=iso&tlng=en}.

\bibitem[Mahdavi-Hezaveh et~al.(2021)Mahdavi-Hezaveh, Dremann, and
  Williams]{mahdavi2021software}
Rezvan Mahdavi-Hezaveh, Jacob Dremann, and Laurie Williams.
\newblock Software development with feature toggles: practices used by
  practitioners.
\newblock \emph{Empirical Software Engineering}, 26\penalty0 (1):\penalty0
  1--33, 2021.

\bibitem[Mann and Whitney(1947)]{mann_test_1947_22}
H.~B. Mann and D.~R. Whitney.
\newblock On a {Test} of {Whether} one of {Two} {Random} {Variables} is
  {Stochastically} {Larger} than the {Other}.
\newblock \emph{The Annals of Mathematical Statistics}, 18\penalty0
  (1):\penalty0 50--60, March 1947.
\newblock ISSN 0003-4851, 2168-8990.
\newblock \doi{10.1214/aoms/1177730491}.
\newblock URL \url{https://projecteuclid.org/euclid.aoms/1177730491}.

\bibitem[Meinicke et~al.(2020{\natexlab{a}})Meinicke, Hoyos, Vasilescu, and
  K{\"a}stner]{meinicke2020capture}
Jens Meinicke, Juan Hoyos, Bogdan Vasilescu, and Christian K{\"a}stner.
\newblock Capture the feature flag: Detecting feature flags in open-source.
\newblock In \emph{International Conference on Mining Software Repositories},
  2020{\natexlab{a}}.

\bibitem[Meinicke et~al.(2020{\natexlab{b}})Meinicke, Wong, Vasilescu, and
  K{\"a}stner]{meinicke2020exploring}
Jens Meinicke, Chu-Pan Wong, Bogdan Vasilescu, and Christian K{\"a}stner.
\newblock Exploring differences and commonalities between feature flags and
  configuration options.
\newblock In \emph{Proc. Int’l Conf. Software Engineering--Software
  Engineering in Practice (ICSE-SEIP). ACM}, 2020{\natexlab{b}}.

\bibitem[Moura and Murta(2018)]{moura_uma_2018_20}
Tayane Moura and Leonardo Murta.
\newblock Uma técnica para a quantificação do esforço de merge.
\newblock In \emph{Proceedings of the 6th Workshop on Software Visualization,
  Evolution and Maintenance}, São Carlos, SP, Brasil, 2018.

\bibitem[Neely and Stolt(2013)]{neely_continuous_2013_8}
S.~Neely and S.~Stolt.
\newblock Continuous {Delivery}? {Easy}! {Just} {Change} {Everything} ({Well},
  {Maybe} {It} {Is} {Not} {That} {Easy}).
\newblock In \emph{2013 {Agile} {Conference}}, pages 121--128, Nashville, TN,
  USA, August 2013.
\newblock \doi{10.1109/AGILE.2013.17}.

\bibitem[Parnin et~al.(2017)Parnin, Helms, Atlee, Boughton, Ghattas, Glover,
  Holman, Micco, Murphy, Savor, et~al.]{parnin2017adages}
Chris Parnin, Eric Helms, Chris Atlee, Harley Boughton, Mark Ghattas, Andy
  Glover, James Holman, John Micco, Brendan Murphy, Tony Savor, et~al.
\newblock The top 10 adages in continuous deployment.
\newblock \emph{IEEE Software}, 34\penalty0 (3):\penalty0 86--95, 2017.

\bibitem[Phillips et~al.(2011)Phillips, Sillito, and
  Walker]{phillips2011branching}
Shaun Phillips, Jonathan Sillito, and Rob Walker.
\newblock Branching and merging: an investigation into current version control
  practices.
\newblock In \emph{Proceedings of the 4th International Workshop on Cooperative
  and Human Aspects of Software Engineering}, pages 9--15, 2011.

\bibitem[Prudêncio et~al.(2012)Prudêncio, Murta, Werner, and
  Cepêda]{prudencio_lock_2012_19}
João~Gustavo Prudêncio, Leonardo Murta, Cláudia Werner, and Rafael Cepêda.
\newblock To lock, or not to lock: {That} is the question.
\newblock \emph{Journal of Systems and Software}, 85\penalty0 (2):\penalty0
  277--289, February 2012.
\newblock ISSN 0164-1212.
\newblock \doi{10.1016/j.jss.2011.04.065}.
\newblock URL
  \url{http://www.sciencedirect.com/science/article/pii/S0164121211001063}.

\bibitem[Rahman et~al.(2016)Rahman, Querel, Rigby, and
  Adams]{rahman_feature_2016_4}
Md~Tajmilur Rahman, Louis-Philippe Querel, Peter~C. Rigby, and Bram Adams.
\newblock Feature toggles: practitioner practices and a case study.
\newblock In \emph{Proceedings of the 13th {International} {Conference} on
  {Mining} {Software} {Repositories}}, pages 201--211. ACM, 2016.

\bibitem[Rahman et~al.(2018)Rahman, Rigby, and Shihab]{rahman_modular_2018_6}
Md~Tajmilur Rahman, Peter~C. Rigby, and Emad Shihab.
\newblock The modular and feature toggle architectures of {Google} {Chrome}.
\newblock \emph{Empirical Software Engineering}, pages 1--28, 2018.

\bibitem[Ramanathan et~al.(2020)Ramanathan, Clapp, Barik, and
  Sridharan]{ramanathan2020piranha}
Murali~Krishna Ramanathan, Lazaro Clapp, Rajkishore Barik, and Manu Sridharan.
\newblock Piranha: reducing feature flag debt at uber.
\newblock In \emph{Proceedings of the ACM/IEEE 42nd International Conference on
  Software Engineering: Software Engineering in Practice}, pages 221--230,
  2020.

\bibitem[Rehn(2012)]{rehn_continuous_2012_9}
Christian Rehn.
\newblock Continuous {Integration}: {Aspects} in {Automation} and
  {Configuration} {Management}.
\newblock Term {Paper}, TU Kaiserslautern, Germany, 2012.

\bibitem[Romano et~al.(2006)Romano, Kromrey, Coraggio, Skowronek, and
  Devine]{romano_exploring_2006_24}
Jeanine Romano, Jeffrey~D. Kromrey, Jesse Coraggio, Jeff Skowronek, and Linda
  Devine.
\newblock Exploring methods for evaluating group differences on the {NSSE} and
  other surveys: {Are} the t-test and {Cohen}’sd indices the most appropriate
  choices.
\newblock In \emph{annual meeting of the {Southern} {Association} for
  {Institutional} {Research}}. Citeseer, 2006.

\bibitem[Sato(2014)]{sato_bliki_2014_11}
Danilo Sato.
\newblock bliki: {CanaryRelease}.
\newblock \url{https://martinfowler.com/bliki/CanaryRelease.html}, June 2014.

\bibitem[Schermann et~al.(2016)Schermann, Cito, Leitner, Zdun, and
  Gall]{schermann_empirical_2016_7}
Gerald Schermann, Jürgen Cito, Philipp Leitner, Uwe Zdun, and Harald Gall.
\newblock An empirical study on principles and practices of continuous delivery
  and deployment.
\newblock \emph{PeerJ Preprints}, 4, March 2016.
\newblock ISSN 2167-9843.
\newblock \doi{10.7287/peerj.preprints.1889v1}.
\newblock URL \url{https://doi.org/10.7287/peerj.preprints.1889v1}.

\bibitem[Shihab et~al.(2012)Shihab, Bird, and Zimmermann]{shihab_effect_2012_1}
Emad Shihab, Christian Bird, and Thomas Zimmermann.
\newblock The effect of branching strategies on software quality.
\newblock In \emph{Proceedings of the 2012 ACM-IEEE International Symposium on
  Empirical Software Engineering and Measurement}, pages 301--310, Lund,
  Sweden, 2012. ACM.
\newblock ISBN 978-1-4503-1056-7.

\bibitem[Sullivan and Feinn(2012)]{sullivan_using_2012_27}
Gail~M. Sullivan and Richard Feinn.
\newblock Using {Effect} {Size}—or {Why} the {P} {Value} {Is} {Not} {Enough}.
\newblock \emph{Journal of Graduate Medical Education}, 4\penalty0
  (3):\penalty0 279--282, September 2012.
\newblock ISSN 1949-8349.
\newblock \doi{10.4300/JGME-D-12-00156.1}.
\newblock URL \url{https://www.ncbi.nlm.nih.gov/pmc/articles/PMC3444174/}.

\bibitem[Warner(2012)]{warner2012applied}
Rebecca~M Warner.
\newblock \emph{Applied statistics: From bivariate through multivariate
  techniques}.
\newblock Sage Publications, 2012.

\bibitem[Wilcoxon(1945)]{wilcoxon_individual_1945_25}
Frank Wilcoxon.
\newblock Individual comparisons by ranking methods.
\newblock \emph{Biometrics bulletin}, 1\penalty0 (6):\penalty0 80--83, 1945.

\end{thebibliography}
\end{document}